\documentclass{iau} 
\usepackage{graphicx}
\usepackage{bm}
\usepackage[english]{babel}
\usepackage[utf8]{inputenc}
\usepackage{natbib}
\usepackage[table]{xcolor}
\graphicspath{{./fig/}{./png/}}
\newcommand{\brac}[1]{\langle #1 \rangle}
\newcommand{\pd}{\partial}

\newcommand{\mean}[1]{\overline{#1}}

\def\Ro{\mbox{\rm Ro}}

\def\Rm{\mbox{\rm Re}_M}

\def\Rm{R_{\rm m}}

\def\Rey{\mbox{\rm Re}}

\def\Rs{R_{\odot}}

\def\urms{u_{\rm rms}}

\def\etat{\eta_{\rm t}}

\title[Global dynamo simulations] %% give here short title %%
{Global simulations of stellar dynamos}

\author[G. Guerrero]   %% give here short author list %%
{G. Guerrero$^1$}

\affiliation{$^1$Physics Department, Universidade Federal de Minas Gerais,
Av. Antonio Carlos, 6627, Belo Horizonte, MG 31270-901, Brazil \\
email: {\tt guerrero@fisica.ufmg.br}}

\pubyear{2020}
\volume{354}  %% insert here IAU Symposium No.
\jname{Solar and Stellar Magnetic Fields: Origins and Manifestations}
\editors{}
\begin{document}

\maketitle

\begin{abstract}
The dynamo mechanism, responsible for the solar magnetic activity, is still an
open problem in astrophysics. Different theories proposed to explain such phenomena
have failed in reproducing the observational properties of the solar magnetism. Thus,
{\it ab-initio} computational modeling of the convective dynamo in a spherical shell 
turns out as the best alternative to tackle this problem. In this work we review
the efforts performed in global simulations over the past decades. Regarding the 
development and sustain of
mean-flows, as well as mean magnetic field, we discuss the points of agreement 
and divergence between the different modeling strategies.  Special attention is given
to the implicit large-eddy simulations performed with the EULAG-MHD code.
%% add here a maximum of 10 keywords, to be taken form the file <Keywords.txt>
\end{abstract}

\firstsection % if your document starts with a section,
              % remove some space above using this command.
\section{Introduction}

The Sun exhibits a large-scale magnetic field which is believed to be driven by a 
dynamo somewhere in its interior.  As a consequence of the dynamo several
processes in the upper layers of the Sun define the solar activity (i.e., the 11-years
sunspot cycle, the solar wind, coronal mass ejections and flares).  It establishes a 
close interaction between the Sun and the earth defining the space weather 
and, perhaps, influencing long term variations of the earth's temperature.  
Besides being fascinating as a physical problem, the astrophysical dynamo is a highly 
relevant area for our society. 

\begin{figure}[htb]
\begin{center}
\includegraphics[width=0.99\textwidth]{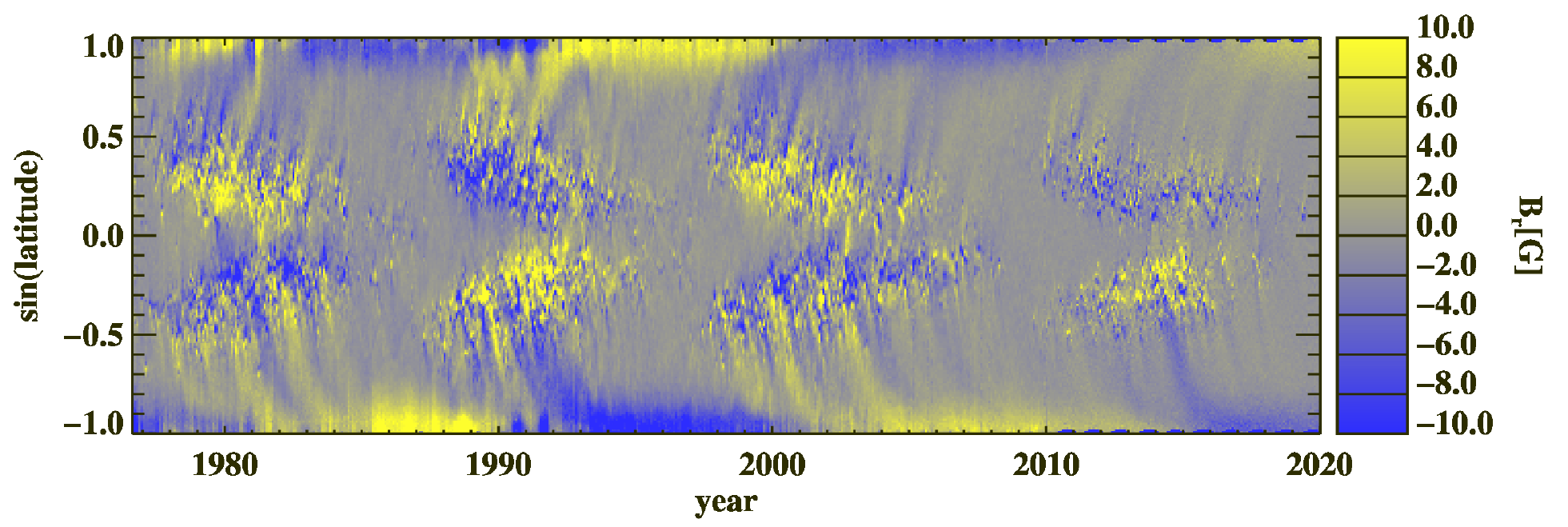}\\
\includegraphics[width=0.42\textwidth]{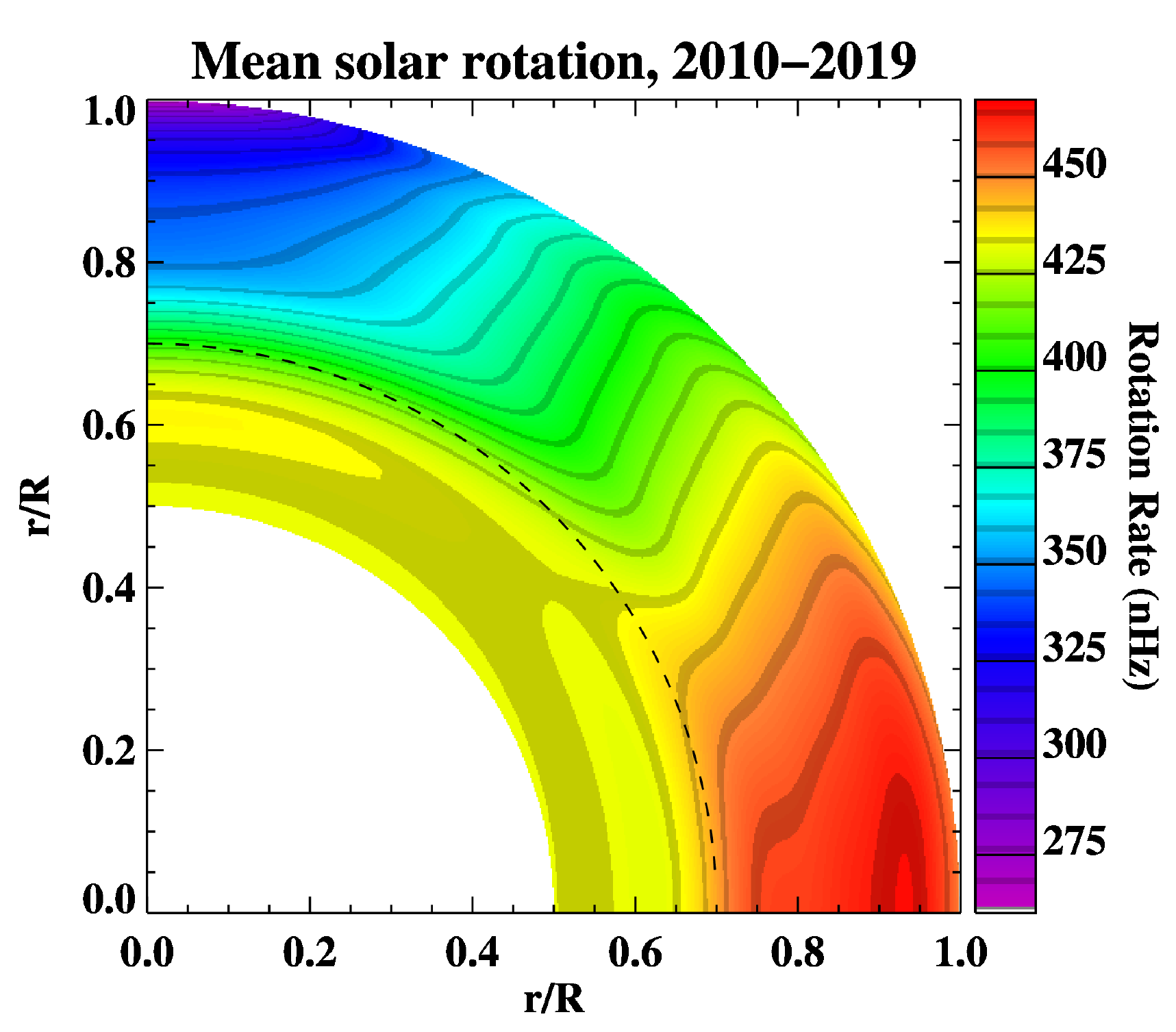}
\includegraphics[width=0.48\textwidth]{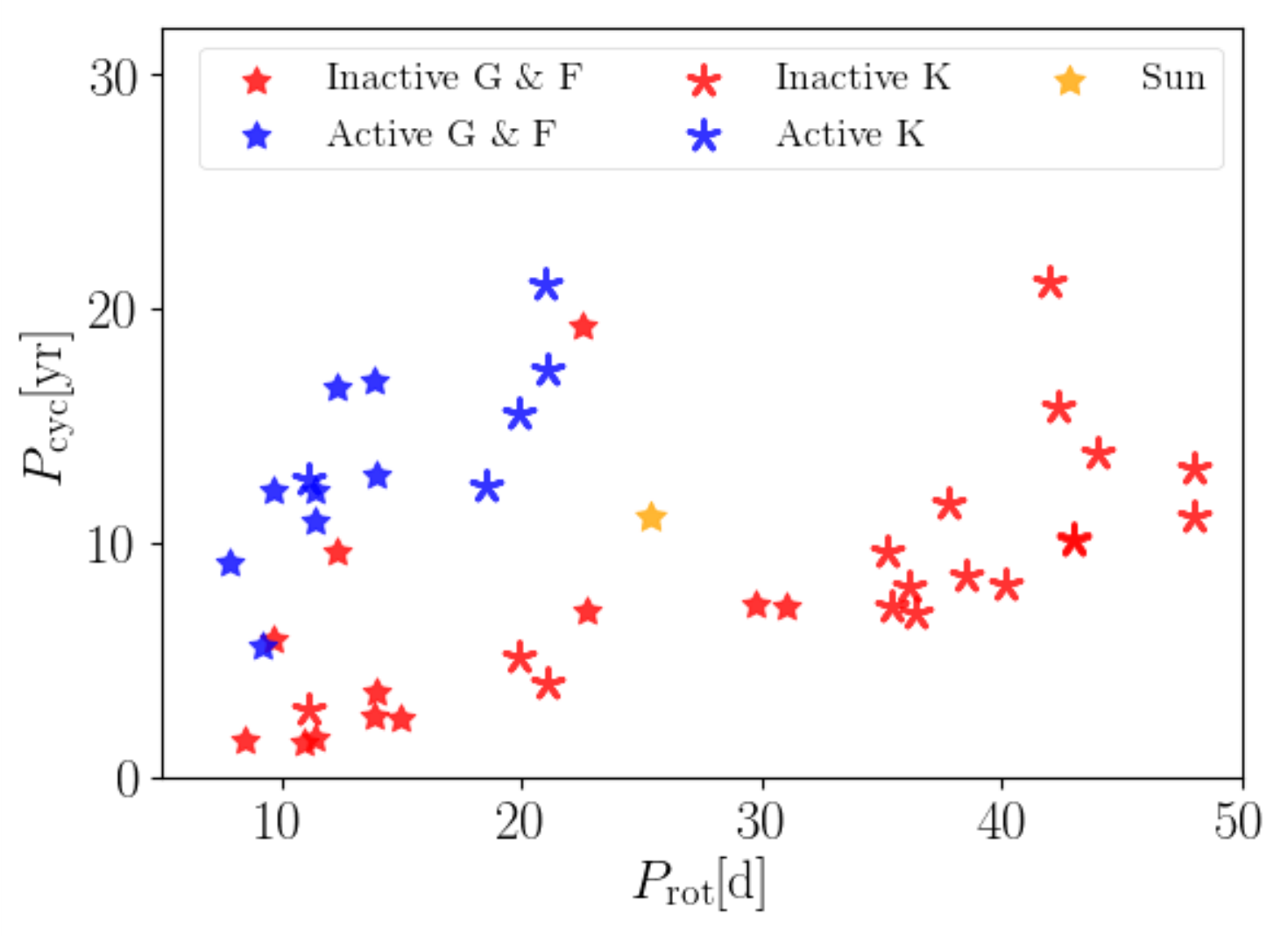}
\caption{Top: time-latitude butterfly diagram (courtesy: A. Kosovichev); bottom-left: solar 
differential rotation (courtesy:  A. Kosovichev); bottom-right: relation between  $P_{\rm rot}$ 
and $P_{\rm cyc}$. The blue and red stars correspond to the observed active and
inactive branches, respectively. The data was taken from \cite{BMM17}.  
\label{fig.1}}
\end{center}
\end{figure}

The main properties of the solar cycle are presented in Fig.~\ref{fig.1}(top panel) and can be
summarized as follow: (1) Sunspots appear in pairs of opposite polarity at latitudes 
of about $30^{\circ}$ and migrate towards the equator, (2) Spots in the northern 
hemisphere have the opposite polarity than their analogs in the southern hemisphere,
(3) When the number of sunspots is maximum, the poloidal field reaches its minimum
and reverses polarity, (4) accordingly, when the poloidal field is maximum, sunspots of 
a new cycle start to appear with opposite polarity in both hemispheres. 

In addition to the solar data, recent observations of magnetic fields
in main sequence stars of types F, G and K have imposed further constrains to the
dynamo mechanism. Observations indicate two important relations
between rotational period of the stars, $P_{\rm rot}$ and the observed magnetic field.
First, the field strength exhibits two regimes:  for fast rotation it  
is independent of $P_{\rm rot}$,  for slow rotation the field amplitude decays with 
$P_{\rm rot}$ with power law dependence \citep[e.g.][]{Wright+11,Vidotto+14}\footnote{Fully convective 
stars follow also this trend.}. The second relation regards the stars that exhibit
activity cycles. A comparison between the magnetic cycle period, $P_{\rm cyc}$,
and $P_{\rm rot}$ shows two regimes of activity, the so called active and 
inactive branches.  For both of them, the longer $P_{\rm rot}$, the longer the
$P_{\rm cyc}$ (Fig~\ref{fig.1}, bottom-right).

Models to understand the solar/stellar dynamo problem are based on the theoretical 
framework known as mean-field dynamo theory \citep{Pa55,SKR66}. It describes 
the generation of the mean magnetic field as the results of the inductive action
of large-scale shear-flows ($\Omega$-effect) and small scale, turbulent, helical motions
and currents ($\alpha$-effect). The properties of the $\Omega$-effect, i.e., the 
solar differential rotation, have been accurately identified by helioseismology 
(Fig.~\ref{fig.1}, bottom-left). Recent asteroseismology results have inferred that similar
rotational shear profiles occur in solar type stars \citep{Benomar+18}. On the other side, the 
values and profiles of the $\alpha$-effect in the solar convection zone are uncertain. 
Some heuristic guesses for these contribution have found some success in reproducing the main
characteristics of the solar cycle, nevertheless the results are ambiguous.  

From the seminal work of \cite{GM81}, three dimensional, first principles, global 
simulations were developed with the goal of unambiguously determine the location and
properties of the dynamo sources. These models have been successful in reproducing
and explaining the mechanisms that sustain the differential rotation 
\citep[DR, see e.g., ][]{MBT06,KMGBC11,GSKM13b,FM15}, yet the magnetism has still some caveats. 
The growing of magnetic field in spherical turbulent convection resembling
the solar case was obtained by \cite{BMT04}. They found magnetic fields
with erratic behaviour without large-scale dynamics.  More recently, \cite{GCS10}
obtained for the first time cyclic dynamo action in a simulation using implicit
sub-grid scale (SGS) modeling with the EULAG-MHD code. The magnetic cycle period of 
their simulation was about $60$ years, yet the migration of the magnetic field did 
not reproduce quite well the solar activity pattern. Later on, other groups have obtained oscillatory 
dynamos \citep[e.g.,][]{KMB12,ABMT15} in spherical shells that consider only the convection zone.
Their results are characterized by short dynamo cycle periods.
\cite{GSDKM16a} compared dynamo models with and without the radial shear layer
at the interface between the radiative and the convective zone, the tachocline.
They found important differences in the process of generating large scale magnetic
fields and noticed that when the tachocline is included, $P_{\rm cyc}$ 
could be of the order of decades. 

More recently the role of rotation was explored for simulations mimicking the
solar interior. At odds with the observations, simulations including the convection 
zone only found that the cycle period decreases as the rotational period increases
\citep{SBCBdN17}. On the other hand, \cite{GZSDKM19} found 
proportionality between $P_{\rm cyc}$ and $P_{\rm rot}$ in simulations including
also a fraction of the convection zone. The results, however,
do not completely fit to the active or inactive branches of activity. 

The numerical resolution used in the simulations 
discussed above is still far from capturing all the relevant scales of the
solar (or stellar) interior. To the date, the simulations with highest resolution were
run for less than $100$ years \citep{Viviani+18,Hotta+16}.  The simulations of
\cite{Hotta+16} have the maximal resolution reported in the literature and were run
for $50$ yr. These simulations show 
reversals of magnetic field and, more importantly, 
are able to develop small scale dynamo action. Thus, the contribution of the small 
scale Maxwell stresses is captured in the simulations. Unfortunately, the characteristics
of the magnetic field, evolution and periodicity, still diverge from the solar case. 
Parametric analyses at this resolution are still prohibitive. For simulations where
the cycle periods are of the order of decades, the evolution time should
be several hundred years for the variables to achieve statistically steady state.
Therefore, large computing resources are needed.

The results described above make clear that in spite of the recent progress
in observations and the development of high resolution simulations,
the solar/stellar dynamo problem is not closed. There are some convergent 
results from which we have learn much about the behavior of stellar interiors.
However, there is not yet a complete theory
to satisfactory explain the details captured by the observations. 
In this work we review the recent advances in {\it ab-initio} global simulations
of solar and stellar dynamos. We discuss the encouraging results as well as the
caveats in the different approaches of global modeling. Even though great progress 
has been done in simulating fully convective stars, we focus here in solar-like stars with
an inner radiative zone and a convective envelope.   

This paper is organized as follows. In the following section we describe the
equations of magnetohydrodynamics (MHD), the framework on which the global
simulations are built, and some characteristics of the simulations setups. 
In \S\ref{s.dnsles} we discuss the direct numerical simulations and the large eddy
simulations  approaches for modeling rotating turbulent convection.  
In \S\ref{s.mf} and in \S\ref{s.bf} we examine the results regarding mean-flow 
and large-scale magnetic field development in stellar interiors. A brief discussion
of the theoretical interpretation of the results is also presented. 
Finally, we draw some conclusions in \S\ref{s.c}.

\section{Magnetohydrodynamics in stellar interiors}

The mathematical formalism describing the dynamics of the plasma in
stellar interiors combines the equations of Navier-Stokes for a magnetized
fluid with the magnetic field induction equation.  The goal is simulating
the motions in convection zones of the Sun and other similar stars. 
With exception of the upper part of the solar convection zone, these motions
are sub-sonic. Thus, the anelastic approximation, which relaxes the Courant
condition imposed from sound waves on the simulations time step, has been
broadly used. Under this approach the MHD equations are the following:
\begin{equation}                                                                                               
{\bm \nabla}\cdot(\rho_s\bm u)=0, \label{equ:cont}
\end{equation}
\begin{equation}
        \frac{D \bm u}{Dt}+ 2{\bm \Omega} \times {\bm u} =  
    -{\bm \nabla}\left(\frac{p'}{\rho_s}\right) + {\bf g}\frac{\Theta'}
     {\Theta_s} + \frac{1}{\mu_0 \rho_s}({\bm B} \cdot \nabla) {\bm B} 
    + \cal{D}_{\bf u} \;, \label{equ:mom} 
\end{equation}
\begin{equation}
        \frac{D \Theta'}{Dt} =  {\cal{D}}_{\Theta} \;, \label{equ:en} 
\end{equation}
\begin{equation}
 \frac{D {\bm B}}{Dt} = ({\bf B}\cdot \nabla) {\bm u} - {\bm B}(\nabla \cdot {\bm u})  + \cal{D}_{\bf B}\;,
 \label{equ:in} 
\end{equation}
\begin{equation}
\nabla \cdot {\bm B} = 0 \; ,
\end{equation}
\noindent
where $D/Dt = \pd/\pd t + \bm{u} \cdot {\bm \nabla}$ is the total
time derivative, ${\bm u}$ is the velocity field in a rotating
frame with ${\bm \Omega}=\Omega_0(\cos\theta,-\sin\theta,0)$,
$p'$ is a pressure perturbation variable that accounts for both the gas
and magnetic pressure,
${\bm B}$ is the magnetic field, and $\Theta'$ is the potential temperature
perturbation with respect to a background state. It is related to the specific 
entropy via $s=c_p \ln\Theta+{\rm const}$;
${\bm g}=GM/r^2 \bm{\hat{e}}_r$
is the gravity acceleration, $G$ and $M$ are the gravitational
constant and the stellar mass, respectively, and $\mu_0$ is the magnetic
permeability. The $\cal{D}$ terms in the equations (\ref{equ:mom})-(\ref{equ:in}) are
dissipative terms which diffuse momentum, heat and magnetic field. Inside stars
the dissipation coefficients may be computed with the Spitzer formula \citep{spitzer62}.
In the upper part of the convection zone there is radiative cooling because
of hydrogen ionization, nevertheless, global simulations do not include this
effect  because of its numerical cost.

\subsection{Simulation domains}

Convective dynamos are modeled in spherical coordinates. Most of the
simulations cover the entire longitudinal and latitudinal extent, i.e., 
$0 \le \varphi < 2 \pi$, and $0 \le \theta \le \pi$, respectively.  The radial 
extent spans for most of the convection zone, $0.71 \Rs \le r \le 0.95 \Rs$, for the
case of the Sun. 
Simulations with the pencil-code consider wedges of different longitudinal extents,
and a latitudinal extent that does not reach the poles to avoid shorter time
scales due to the spherical grid \citep[e.g.,][]{KMB12}. However, since it solves fully 
compressible equations, the radial extent in simulations with this code reaches
up to $1 \Rs$. \cite{BMT11,brun+17}, with the ASH code, and  \cite{GSKM13b}
with the EULAG-MHD code have performed hydrodynamic (HD) simulations including a 
stable stratified layer.  
\cite{MYK13} with a code based on the Yin-Yang grid and second order finite 
differences, as well as \cite{GCS10,GSDKM16a}, with the 
EULAG-MHD code,  performed dynamo simulations which also include a fraction of 
the radiative zone.

\subsection{Direct numerical simulations (DNS) and large eddy simulations (LES)}
\label{s.dnsles}

The dynamo phenomena is a MHD problem associated to rotating turbulent 
convection, i.e., the Reynolds number ($\Rey=\urms L/\nu$, where $\urms$ is the
turbulent velocity,  $L$ a characteristic 
length scale of the system, and $\nu$ is the kinematic viscosity of the plasma) 
and the magnetic Reynolds number ($\Rm=\urms L/\eta$, where $\eta$ is the
magnetic diffusivity) have large values.  In the solar convection zone both quantities
exceed $10^6$.  Numerical simulations capturing all the relevant scales of a turbulent 
flow are called direct numerical simulations (DNS). 
Since the Reynolds numbers give a rough measurement of the range of scales, between 
the advective and the dissipative processes, it can be estimated that the number of 
grid points necessary to resolve all the contributing scales must be 
of the order of $N \sim \Rey^{9/4}$. Thus, DNS simulations reproducing 
the dynamics of the Sun's interior are still prohibitive for the current 
supercomputers (e.g., the simulations of \cite{Hotta+16} reach a maximum Reynolds 
number of about $7000$). Thus, most simulations use values of the dissipation
coefficients much larger than the molecular ones and consistent with the turbulent 
rates of dissipation. 

Other alternatives have been developed over the years to simulate turbulent flows.
These theories aim to mimic the contribution of the non-resolved motions by 
adding a sub-grid scale (SGS) terms in the prognostic equations. This allows to 
model high $\Rey$ flows with less expensive simulations. Since the computations
in this approach resolve only the scales of relatively large structures they are
called large-eddy simulations. In most of the SGS models the contribution 
of the non-resolved scales is proportional to the strain tensor of the large-scale
flows \citep[]{Sma63,GPMC91}. To test the accuracy of this kind of modeling, LES are 
compared, when possible,  with experiments or with high resolution DNS. For instance,
the incompressible simulations with forced turbulence carried out at a resolution
of $4096^3$ mesh points by \cite{Kaneda03} were compared to the Smagorinsky and the
hypervisocity SGS models by \cite{HB04,HB06}. The results showed a good agreement 
between the three cases in the turbulent inertial range. 

More recently, a new class of numerical schemes have obtained results compatible 
with LES simulations without the need of a SGS model. These schemes are based on 
non-oscillatory finite volume (NFV) approximations and the strategy is called 
implicit large-eddy simulation \citep[ILES, see e.g.,][]{Grinstein+07}. There is not yet a turbulence 
theory justifying the basis of the ILES approach, nevertheless, \cite{MR02} presented
a solid rationale for their use by studying the Burger's equation. More recently, 
\cite{Margolin2019}  showed that the terms resulting 
from the NFV formulation might indeed represent physical phenomena. 

ILES modeling of the solar interior were performed during the last decade with
the EULAG-MHD code.  These attempts are summarized in the 
sections below, where we point out the success and caveats of the obtained results. 

\subsection{Global solar/stellar simulations with the EULAG-MHD code}
\label{s.deu}

EULAG-MHD \citep{SC13} is an extension of the hydrodynamic model EULAG 
predominantly used in atmospheric and climate research \citep{PSW08}. Its solver is  
adapted to simulate the turbulent subsonic flows found in the majority of the 
solar interior.  EULAG-MHD is powered by a non-oscillatory forward-in-time MPDATA method 
(multidimensional positive definite advection transport algorithm; see \citet{S06} 
for an overview). This is a nonlinear, second-order-accurate iterative implementation 
of the elementary first-order-accurate flux-form upwind scheme. 

\cite{ES02} performed the first HD ILES simulations of the solar convection zone. 
They were able to create solar-like differential rotation profiles by modeling 
turbulent convection rotating at the solar rotation rate. MHD simulations 
including the radiative zone were performed by \cite{GCS10}. They found, for the
first time, cyclic reversals of magnetic field occurring 
at the tachocline. In their work, and subsequent simulations with EULAG-MHD, the
heat and radiative transfer terms, ${\cal D}_{\theta}$,  are replaced by the simple 
forcing and cooling  parametrization,
\begin{equation}
\frac{D \Theta'}{Dt} = -{\bm u}\cdot {\bm \nabla}\Theta_{\rm amb} -\frac{\Theta'}{\tau}\;. \label{equ.en2}
\end{equation}

\begin{figure}[htb]
\begin{center}
\includegraphics[width=0.5\textwidth]{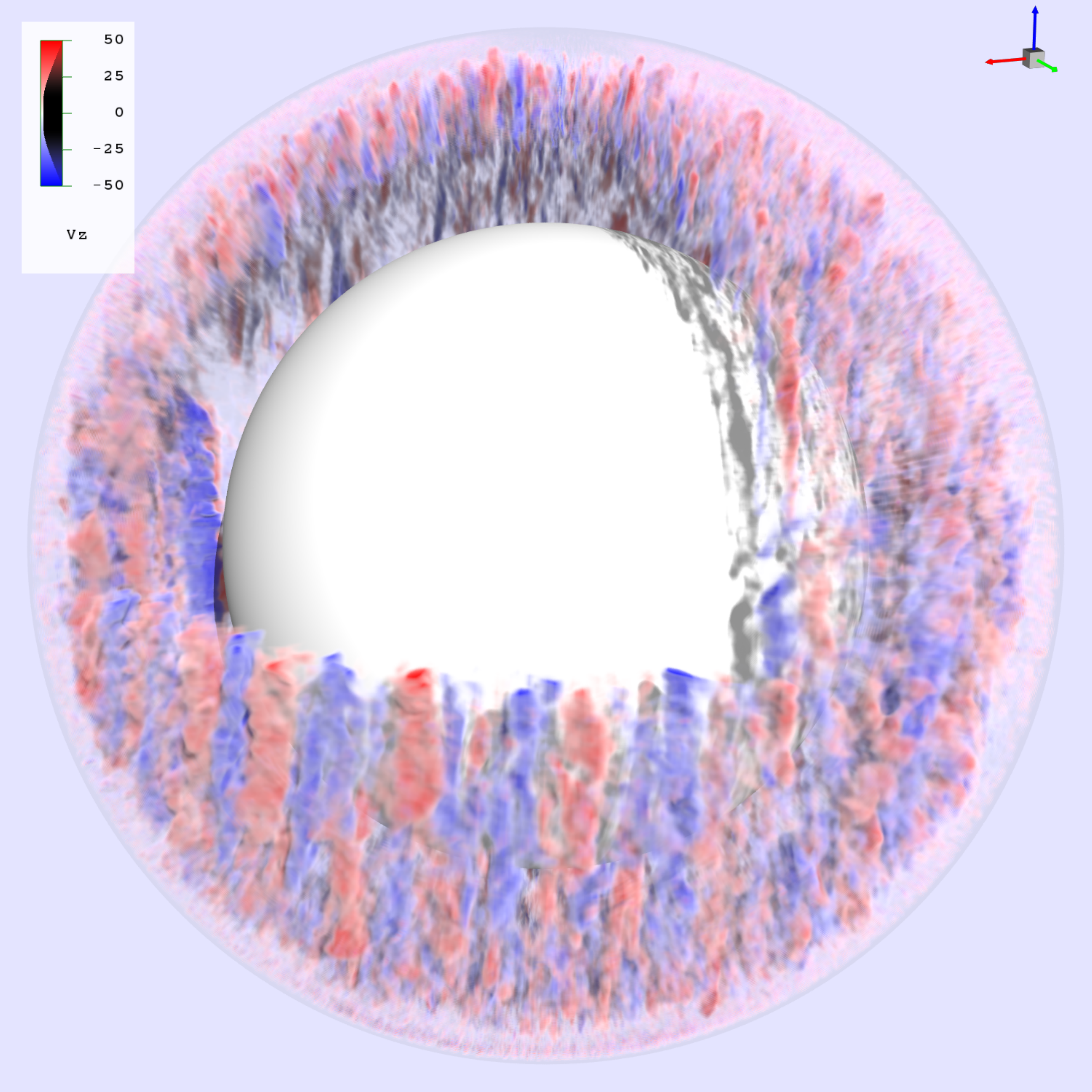}
\caption{Radial velocity component of a characteristic EULAG-MHD simulation of
rotating convection. The elongated convective rolls are the so-called banana
cells, typical from rotationally constrained motions. 
\label{fig.vz}}
\end{center}
\end{figure}

In this form, the Newtonian cooling (second term on the RHS) relaxes  perturbations 
of potential temperature towards an axisymmetric ambient state, $\Theta_{\rm amb}$, in a
time scale, $\tau$ \citep[see][and references therein]{HS94,Cossette+17}. Under this scheme, convection
is driven by these perturbations whenever the ambient is super-adiabatic, and
relaxes towards a statistically steady state on timescales much shorter than the ones 
needed by the transport of heat due to radiation or diffusion. In addition, it simplifies
the energy boundary conditions because only the radial derivative of the radial
convective flux has to be specified \citep{SC13}. As it will be discussed below, the choice of 
the ambient state is pivotal to define not only the strength of the convective
motions, but also the frequency of inertial gravity waves in stable stratified
atmospheres.  Suitable ambient states may be obtained for the Sun and 
other stars from evolutionary codes. The viscosity and magnetic diffusivity
terms are also dropped out from Eqs.~(\ref{equ:mom} and \ref{equ:in}). Thus,  the 
only dissipation in the system is due to the ILES numerical scheme, MPDATA. 
Figure~\ref{fig.vz} depicts the vertical velocity of a characteristic 
EULAG-MHD simulation with a rotational period of $28$ days.  The banana cell
structures, typical of rotationally constrained convection, are evident in
the snapshot. 

Although this approach has been instrumental to obtain mean-flows and cyclic
dynamos resembling the observations, it still has some issues that must
be addressed. For instance,  \cite{GSKM13b} used EULAG-MHD to study how mean-flows 
develop and sustain in simulated stars rotating at different rates. 
Their reference simulations have $128\times 64\times 64$ grid points in the $\varphi$, 
$\theta$ and $r$ directions, respectively. Rotating at the solar rotation rate
it develops solar-like DR.  When the resolution is doubled or 
increased four fold, the resulting DR exhibits different patterns. Thus, 
understanding the interplay between a scale independent Newtonian cooling and 
the implicit numerical dissipation (depending on numerical resolution) is 
necessary. 

\section{Mean flows}
\label{s.mf}

Differential rotation and meridional circulation (MC) are stellar large
scale motions in the longitudinal and the meridional directions, respectively.
Thanks to helioseismology, the differential rotation in the Sun has been measured 
with great accuracy \citep{SCHOU98}.  The profile shown in Fig.~\ref{fig.1} 
(bottom-left panel) 
is a long term average of the angular velocity. Its main characteristics are: fast
equator, slow poles with conical isocontours.  Latitudinal differential rotation 
goes from the surface down to the tachocline at $\sim 0.71 \Rs$, from where the
rotation is rigid. At the surface there is a thin layer of negative shear called
the near-surface shear layer (NSSL). The solar DR suffers periodic speeds-up and 
slows-down called torsional oscillations \citep[e.g.,][]{HL80,KS97,AB01,V+02}.

The meridional circulation, observed
in the solar photosphere with different techniques, shows poleward flows on each
hemisphere. Its amplitude of, about $20$ m s$^{-1}$, also fluctuates with the solar 
cycle. The helioseismic signal of the 
meridional flow from the solar interior is weak. For that reason, despite that
few attempts have been conducted,   
its profile has not yet been accurately measured.  \cite{zbkdh13,str13,boning+17}
inferred meridional circulation patterns with multiple radial cells. 
\cite{liang+18} reported evidences for single cell circulation.

Global simulations aim to reproduce the observed properties of these mean-flows.
However, explaining their peculiarities is still challenging. 
Simulations from diverse numerical techniques and codes have consistently found
a transition from anti-solar DR  (with slow equator and fast poles)  
for slow rotating models, to solar-like DR for
rotationally constrained turbulent convection \citep{gilman77,KMGBC11,GSKM13b,GYMRW14}. 
This transition seems to occur when the Rossby number, measuring the ratio
between the rotational period and the convective turnover time 
($\Ro = P_{\rm rot}/\tau_c$), is about one\footnote{Note, however, that there are
several definitions of the Rossby number and this value might change.}.  
This point marks the transition
between strong and fast convection and slow convective motions influenced by 
the Coriolis force.  Self consistently developed tachoclines have been found in 
recent simulations  \citep[e.g.,][]{GSKM13b,brun+17}. Similarly, the formation of 
the NSSL has been 
studied in recent papers \citep{MHBT19}. 

From the simulation results it is possible to explore how this flows are sustained.
Differential rotation may be explained from the distribution of angular momentum 
${\cal L} = \rho \varpi \mean{u}_{\varphi}$, where $\mean{u}_{\varphi}$ is the mean, 
longitudinally averaged, zonal flow and  $\varpi = r\sin\theta$ is the lever
arm.   In the HD case its evolution equation is
\begin{equation}
 \frac{\partial {\cal L}}{\partial t}
     =- \nabla\cdot \left( \varpi \left[ 
      \rho (\mean{u}_{\varphi}+\varpi \Omega_0)\mean{\bm u}_{\rm m} + 
      \rho \mean{u_{\varphi}' {\bm u}'_{\rm m}} 
%      -\frac{1}{\mu_0} \mean{B}_{\phi}\mean{{\bm B}}_{\rm p}
%      -\frac{1}{\mu_0} \mean{b_{\phi}'{\bm b}_{\rm p}'} 
      \right]  \right ),
      \label{eq.amb} 
\end{equation}
where, $\mean{\bm u}_{\rm m}$, ${\bm u}'_{\rm m}$ 
are the mean and turbulent meridional ($r$ and $\theta$)
components of the velocity field, respectively. All these terms can be computed
from the global simulations.   For solar-like differential
rotation, the second term on the RHS, namely the Reynolds stresses, 
dominate over the meridional circulation terms, first term on the RHS.  
A mostly positive latitudinal Reynolds stress
component, $\rho \varpi \mean{{u}_{\phi}' {u}'_{\theta}}$, transporting 
angular momentum towards the equator,   is a robust 
feature of global simulations of the Sun  \citep{BT02,GSKM13b,FM15, PMCG16}.
This result is in agreement with the $\Lambda$-effect theory
\citep{rudiger89} where the Reynold stresses are parametrized
as $R_{ij} = R_{ij}^{\Lambda} + R_{ij}^{\nu}$. Here, the first and
second terms correspond to non-diffusive and diffusive parts of the
$\Lambda$-effect, respectively.  Positive values of the non diffusive 
part are required to sustain solar-like DR \citep{kit13}. 

For anti-solar DR the meridional circulation terms dominate, advecting
angular momentum towards higher latitudes. \cite{FM15} argue that this 
meridional flow is driven by the transport of angular momentum through the 
so-called gyroscopic pumping mechanism which results from 
Eq.~(\ref{eq.amb}) and considering steady state ($\partial {\cal L}/\partial t =0 $). 
Under this scenario, negative
values of $- \nabla \cdot \varpi \rho \mean{u_{\varphi}' {\bm u}'_{\rm m}}$
will induce a strong meridional flow away from the rotation axis at lower
latitudes. Because of the closed boundary conditions, this flow transports angular 
momentum to the high latitudes increasing the angular velocity. At the
equator it will result in a differential rotation that decreases away
from the rotation axis \citep{MH11}.  

This description is completed with the equation
of the thermal wind balance (TWB) which in its HD form reads,
\begin{equation}
\varpi \frac{\partial \mean{\Omega}^2}{\partial z} = 
\frac{g}{r}\frac{\partial}{\partial \theta}\left(\frac{\Theta^{\prime}}{\Theta_0} \right)\;,
\label{eq.twb}
\end{equation}

\noindent
where $z$ is the height in cylindrical coordinates. If the DR has
isocontours which are not aligned with the rotation axis, the term on the
LHS implies gyroscopic pumping.  The term on the RHS is the baroclinicity 
of the system. Models with anti-solar differential rotation result in 
warm equator and cold poles, favoring then a counterclockwise meridional 
in the northern hemisphere. 
cell \citep{FM15}. The same, gyroscopic pumping mechanism might explain
the, more complex, multi cell pattern of MC typical of simulations with 
solar-like differential rotation \citep{GSDKM16b}. 

An alternative explanation to the gyroscopic pumping dominance in
defining the meridional circulations comes from the $\Lambda$-effect
theory.  Mean-field models under this approximation explain MC from 
departures of the TWB (Eq.~\ref{eq.twb}).  Since the TWB is likely lost
at the boundaries of the convection zone,  viscous forces might drive the
meridional flows \citep{kit13}.  In contrast with global simulations where
all the driving forces arise naturally, in $\Lambda$-effect mean-field simulations
these forces are parametrized by theoretical turbulence models. Thus, including this 
viscous force in the equation for the vorticity 
results in single circulation cells for both fast and slow rotating 
stars \citep{KKC14}.  This disagrees with the results of HD
global simulations where single cell circulation is obtained for slow rotation
and multiple cells show up in fast rotation simulations. (See, however, \cite{PK18} 
for $\Lambda$-effect mean-field simulations with double cell MC.)

\begin{figure}[htb]
\begin{center}
\includegraphics[width=0.45\textwidth]{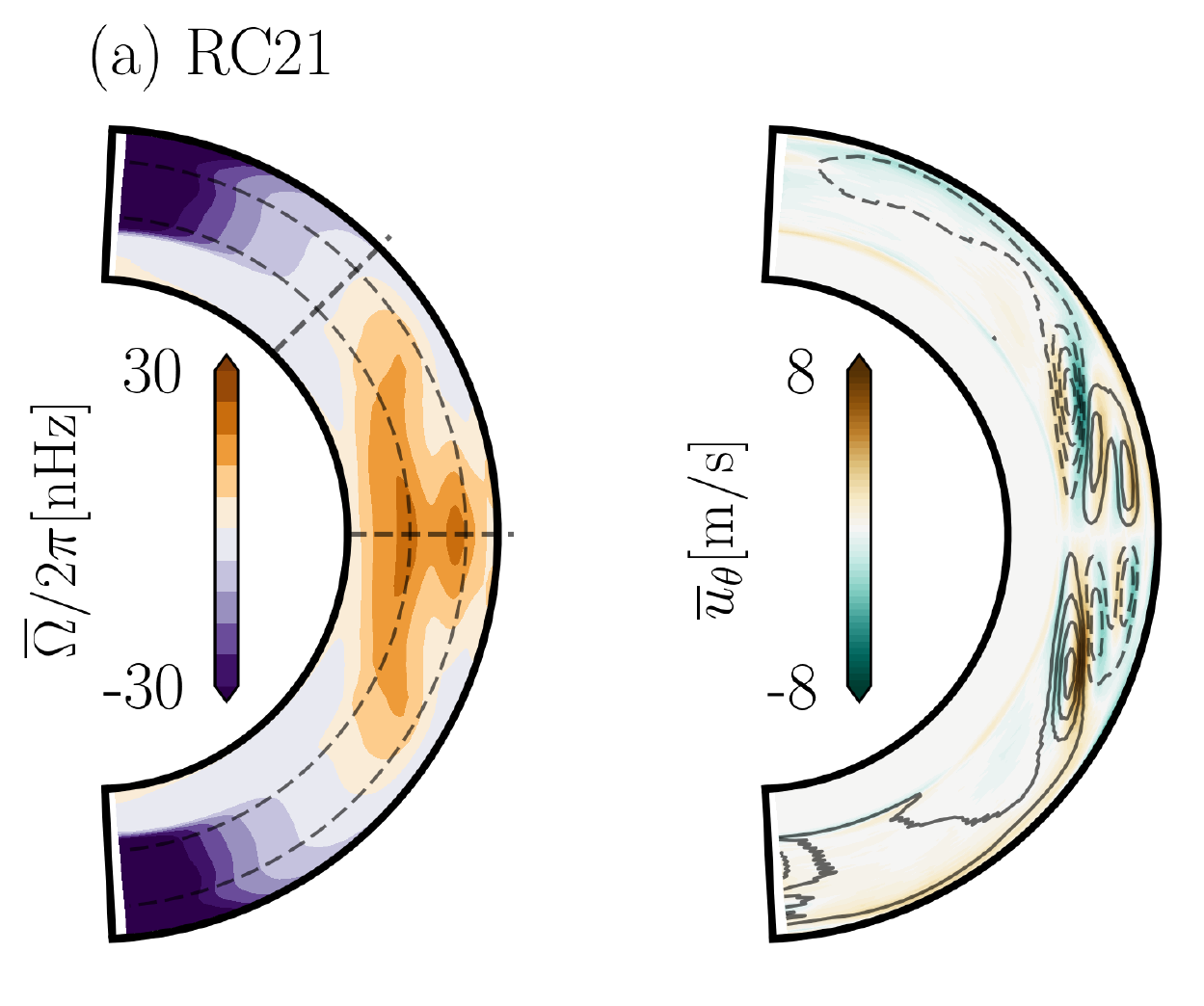}
\includegraphics[width=0.45\textwidth]{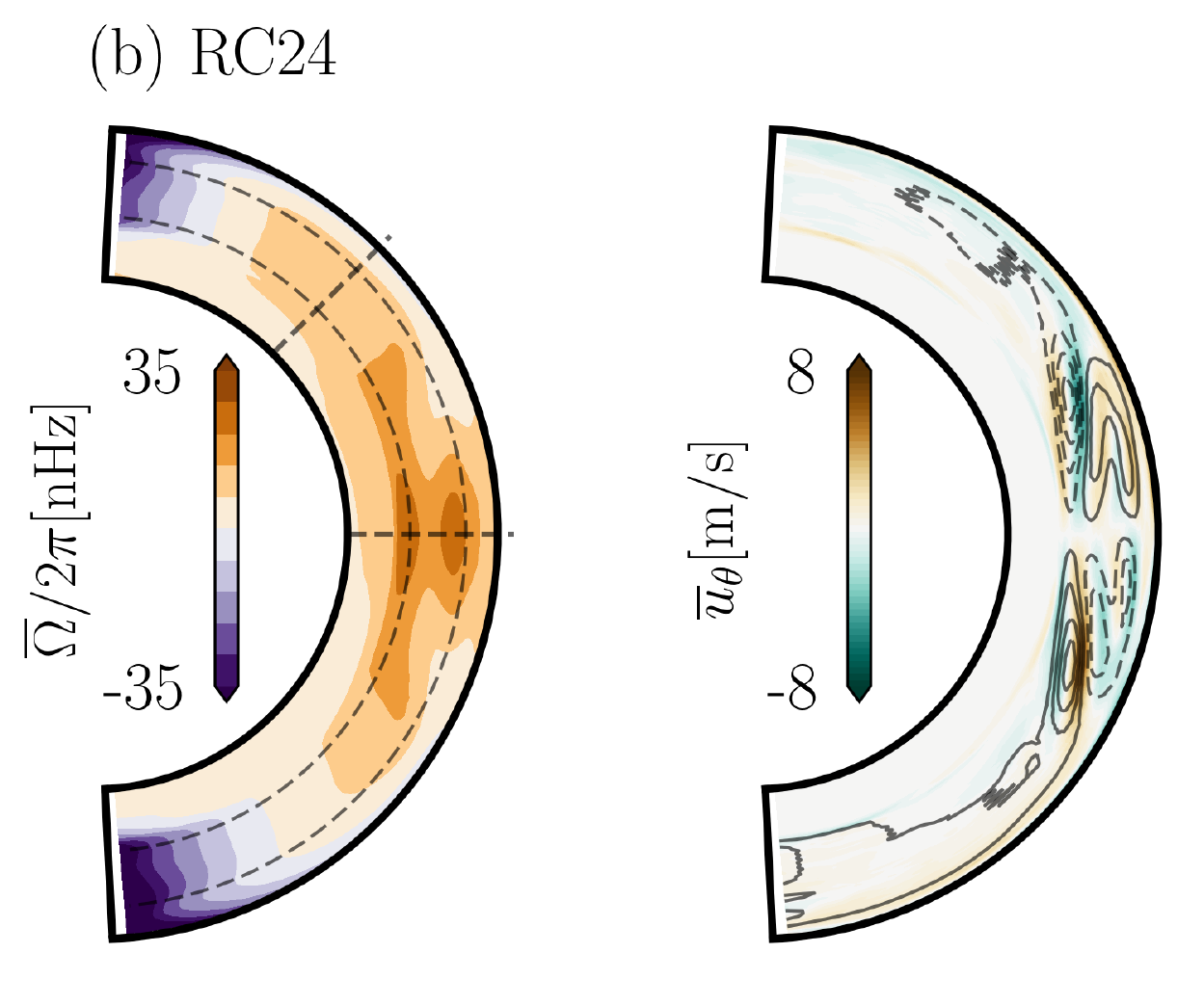}\\
\includegraphics[width=0.45\textwidth]{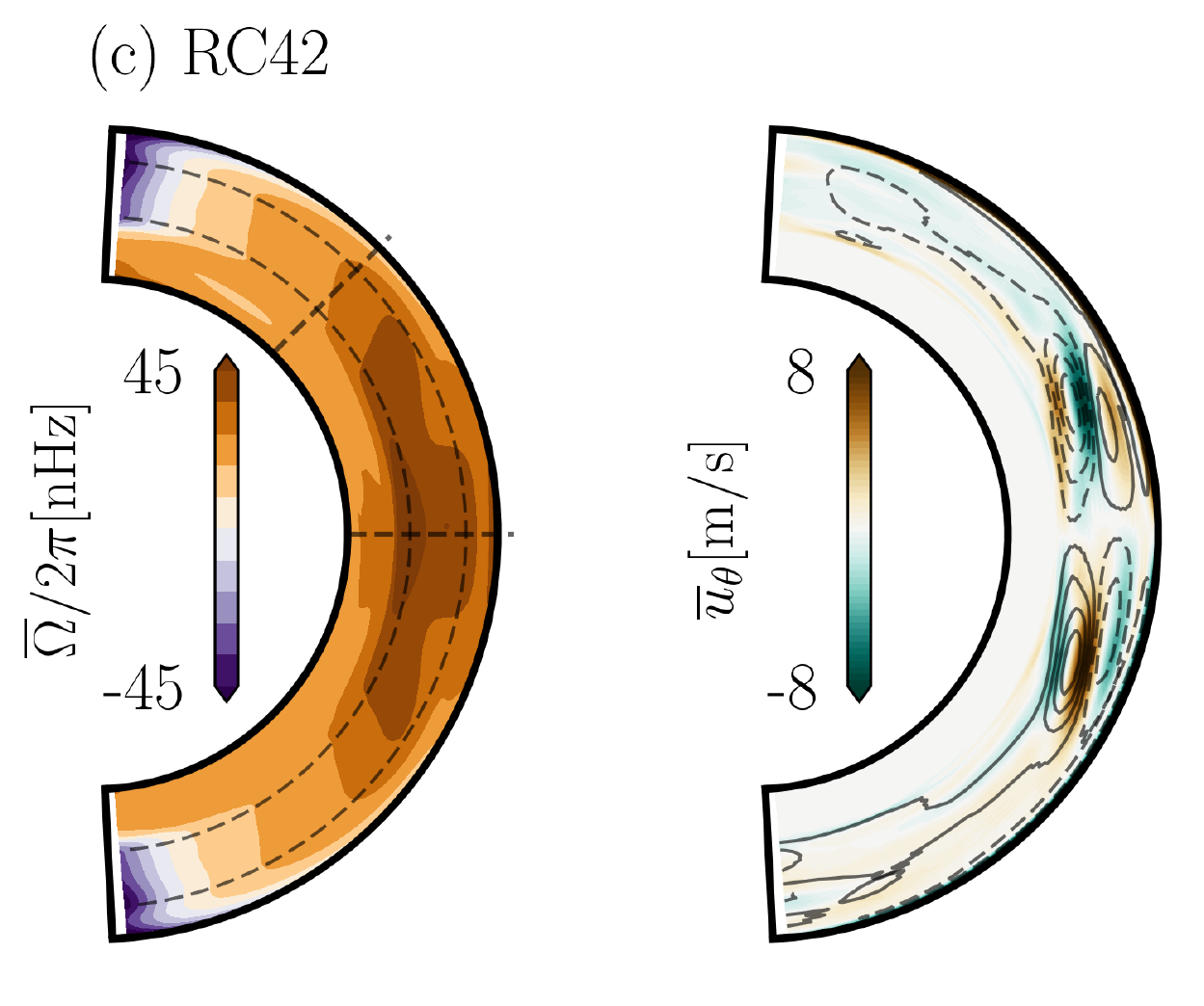}
\includegraphics[width=0.45\textwidth]{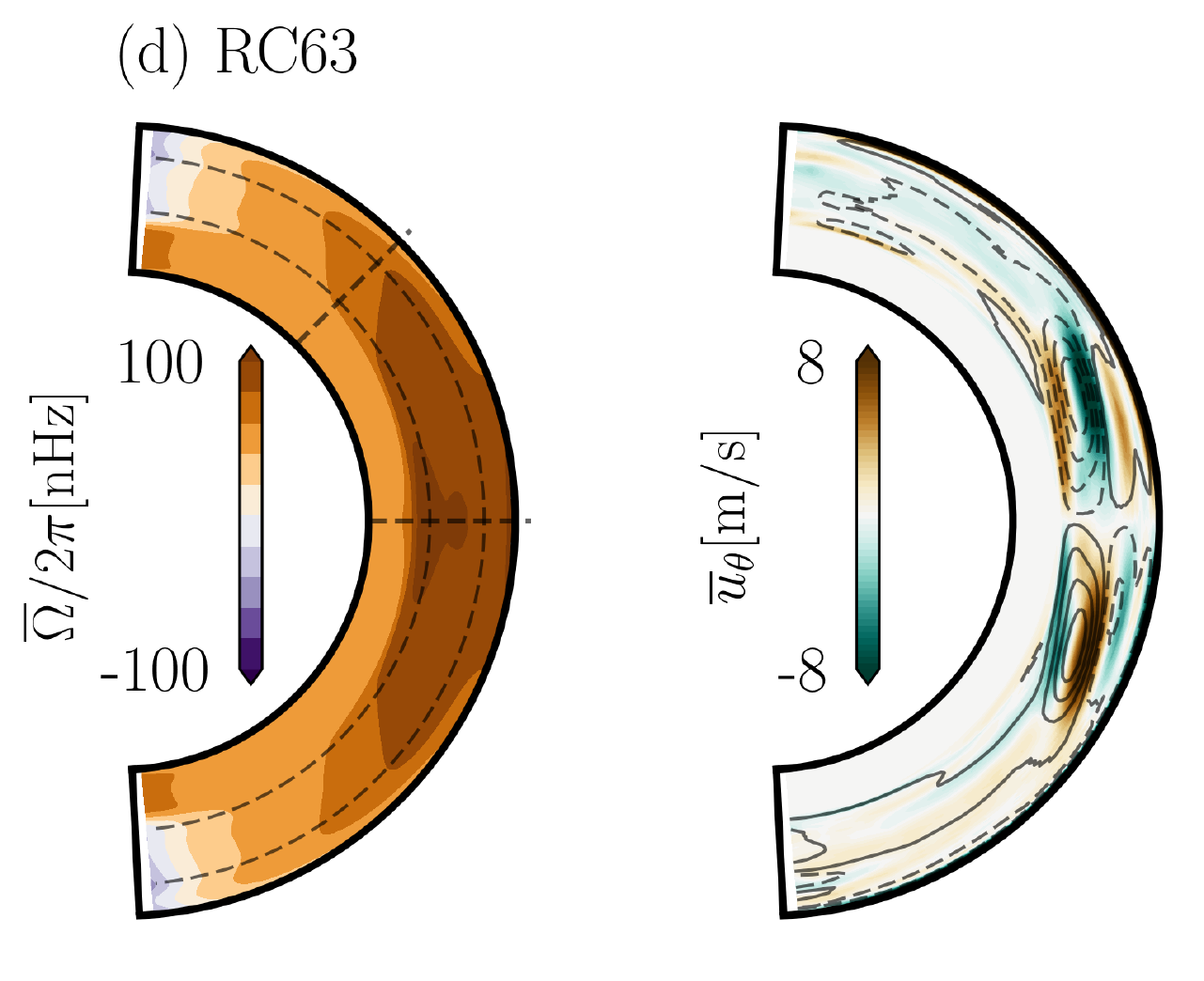}
\caption{Differential rotation in the inertial frame, $\mean{\Omega} - \Omega_0$, 
and meridional circulation of characteristic simulations of
\cite{GZSDKM19}.  In the differential rotation panels the color contours show regions
of iso-rotation. In the meridional circulation panels the continuous (dashed) lines
represent clockwise (counterclockwise) circulation. The color contours show 
the mean latitudinal velocity, $\mean{u}_{\theta}$. Adapted from \cite{GZSDKM19}.
\label{fig.drmc}}
\end{center}
\end{figure}

In the MHD case,  simulations have shown that the magnetic field quenches the 
convective flux affecting 
the distribution of the Reynolds stresses and, therefore, of angular momentum. 
It allows for slow rotating models to develop
solar-like DR \citep{FF14,KKBOP15,GZSDKM19}.  The HD and MHD balance of angular 
momentum and thermal wind for solar-like DR are discussed in detail by \cite{PMCG16}.
The transition from solar-like to anti-solar DR for HD and MHD simulations was 
explored by \cite{KKBOP15}. It was found that the transition is smooth 
for dynamo simulations and occurs at larger values of the 
Rossby number.  In the EULAG-MHD simulations presented in 
\cite{GZSDKM19} this transition is not observed for Rossby numbers up to 
$\sim3$ ($63$ days rotational period). This can be seen in Fig.~\ref{fig.drmc} 
depicting the DR (left of each panel) and MC (right) of some simulations
in that publication. Note also that in all the cases the MC has multiple 
cells in radius at lower latitudes. 

Differential rotation has been observed in other stars by different means
including tracking magnetic features, photometric variability and asteroseismology 
\citep[e.g.,][]{RG15,kovari+17,Benomar+18}. The results indicate 
DR consistent with faster equator for a large number of observed stars. 
Anti-solar DR has not been yet confirmed. \cite{RG15,kovari+17} found
a shear parameter, $\alpha^*=\Delta\Omega/\Omega_{\rm eq}$, increasing with 
the rotational period of the stars with values between $10^{-3}$ and $0.5$.
The recent asteroseismology 
analysis of \cite{Benomar+18} does not show a clear defined trend and found 
shear values larger than unity.  In the Sun $\Delta\Omega/\Omega_{\rm eq}$, with
$\Delta \Omega = \Omega_{\rm eq} - \Omega_{45}$, is about $0.1$.

\begin{figure}[htb]
\begin{center}
\includegraphics[width=0.85\textwidth]{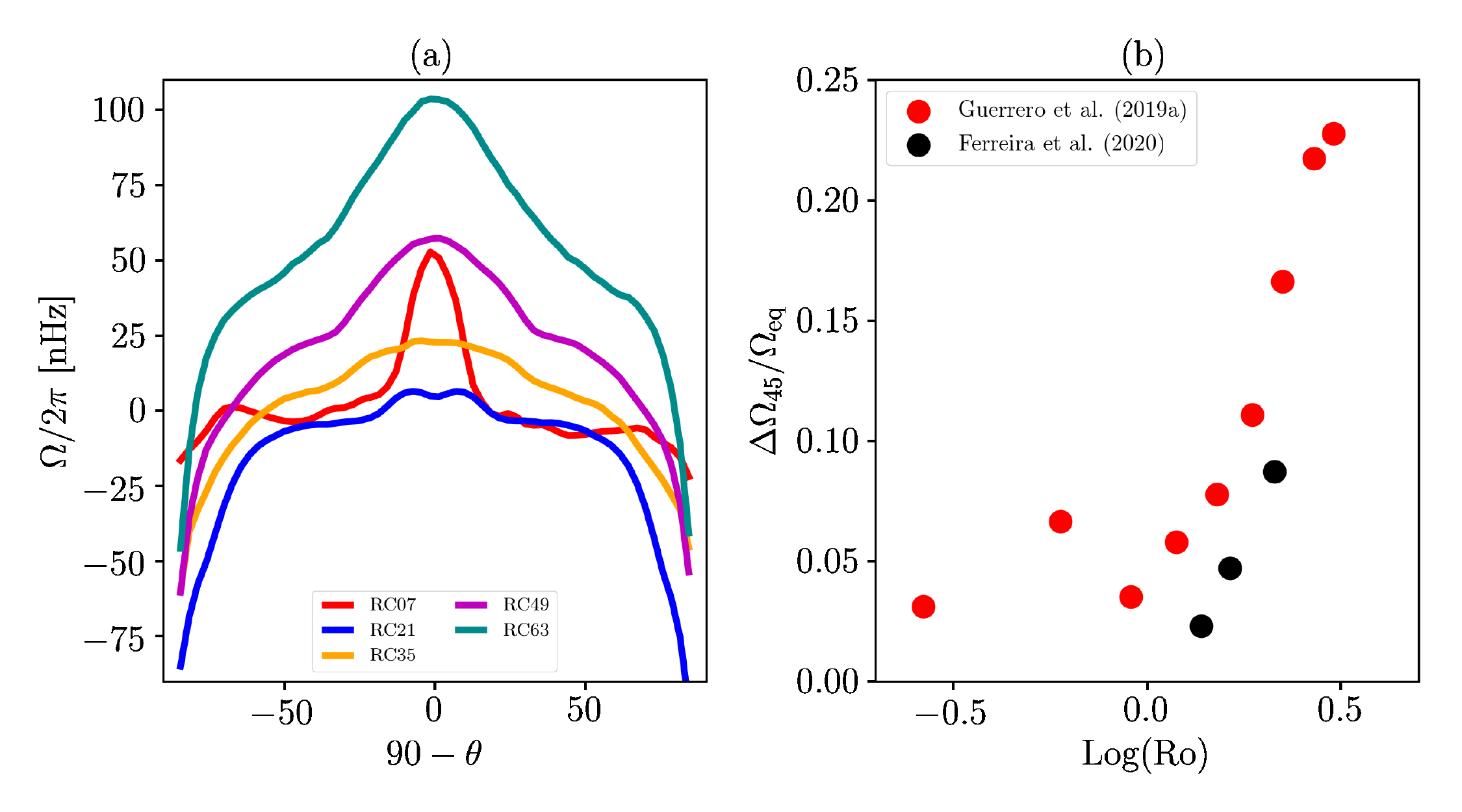}
\caption{(a) Latitudinal differential rotation in the inertial frame, $\mean{\Omega} - \Omega_0$, 
of the simulations presented in \cite{GZSDKM19}; (b) relative shear, $\Delta \Omega /\Omega_{\rm eq}$, 
where $\Delta \Omega = \Omega_{\rm eq} - \Omega_{45}$, as a function of the Rossby number
(in logarithmic scale). The red and black dots correspond to the simulations of \cite{GZSDKM19}
and \cite{ferreira+20}, respectively.
\label{fig.os}}
\end{center}
\end{figure}
Most of the observational findings on DR estimate the shear parameter
by assuming a solar-like latitudinal variation, i.e., $\sim \cos^2\theta$. This, however, 
might not be the case.  Figure~\ref{fig.os} shows the results of the global 
simulations presented in \cite{GZSDKM19}.  The LHS panel shows the latitudinal
profile of $\mean{\Omega}$ in the inertial frame.  It is evident that 
the latitudinal shear has different profiles for different rotational periods. 
The fast rotating simulations show strong variation concentrated at the equator and
the poles. Intermediate cases have smooth profiles while slow rotating
simulations develop a strong shear between the equator and the poles. 
The parameter $\alpha^*$, measuring the shear between the equator and
$45^{\circ}$ \citep[to compare with astroseismic results of][]{Benomar+18}, 
as a function of the Rossby number is presented on the RHS of Fig.~\ref{fig.os}. 
The shear evidently increases with the increase of $\Ro$ (decreasing rotation).
The black dots, corresponding to simulations of the star HD43587 
presented in \cite{ferreira+20},  follow a similar trend yet with
smaller values of $\alpha^*$.  If the increase of the shear with the rotational
period is confirmed by further observations, the physics behind this counterintuitive 
result must be elucidated through global models.

\section{Large-scale magnetic field}
\label{s.bf}

\subsection{The solar dynamo}

For the reasons discussed in previous sections, performing global MHD simulation
of the solar dynamo, including all the features of the convection zone, and  
all the relevant scales involved in the processes, is still unfeasible.  
Nevertheless, in view of the ambiguities of the mean-field simulations, 
global MHD modeling is perhaps the most promising method to understand
the dynamo mechanism in stellar interiors. Several attempts have been done 
over the last decades with the aim of capturing some relevant  physics.  
In the solar context, \cite{GM81,gilman83} presented pseudo-spectral 
simulations of rotating convection in the Boussinesq approximation. They were
able to obtain dynamo action and even magnetic field reversals. The anelastic
approximation was used in the early works of \cite{glatz84,glatz85a,glatz85b}. 
With the advent of parallel supercomputer architectures, simulations
with improved resolutions were performed  by \cite{BMT04}  with the ASH code. 
They obtained amplification of the magnetic field without a large-scale dynamo.

Large-scale organized magnetic fields were found in the simulations performed
by \cite{BMBT06}. They achieve mean-field dynamo action by forcing a shear 
layer, i.e., mimicking the tachocline,  below the convection zone. Nevertheless,
\cite{BBBMT10} showed that the formation of organized global structures do 
not need strong tachocline shear but only rapidly rotating turbulent convection.  
Their simulations, rotating  $\sim 3$ times
faster than the Sun, developed steady antisymmetric magnetic wreaths.

Periodic magnetic fields were first reported by \cite{GCS10} with the ILES 
formulation of the EULAG-MHD code. The lower values of the numerical viscosity
due to the implicit SGS scheme, together with the energy description presented 
in \S\ref{equ.en2} led to a self-consistent development of the tachocline and
a solar-like differential rotation.  The resulting dynamo mode does not 
show, however, clear latitudinal migration and reverses with a cycle period
of 30 yrs. 

Fully compressible simulations of spherical wedges performed with the 
pencil-code also found cyclic dynamo solutions \citep{KKBMT10}. Also,
\cite{KMB12} reported solutions where the toroidal magnetic field migrates
towards the equator resembling the solar observations. 
Considering slope-limited diffusion for diffusing momentum, and 
eddy values for the heat transport coefficients and the magnetic 
diffusivity, \cite{ABMT15} obtained periodic field reversals with the ASH code. 
These cases simulate only in the convection zone and the 
cycle periods are between 3 to 6 yrs.

\cite{MYK13}, with a code based on second order finite differences and using
the Yin-Yang grid explored the role of penetrative convection in models with
and without a fraction of the radiative interior.  They found that larger magnetic
fields develop when the stable layer is present. Nevertheless, there is not a 
well defined tachocline in their results.  \cite{GSDKM16a} performed the
same comparison with EULAG-MHD and explored simulations with rotation rates
of $2 \Omega_{\odot}$, $\Omega_{\odot}$ and  $\Omega_{\odot}/2$. 
The results showed striking differences
between the two cases. While the resulting latitudinal shear was roughly
the same,  the presence of the radial shear at the tachoclines
led to a different behavior of the strong magnetic field developed at 
and below these shear regions. As a consequence, the cycle period, of 
about 2 yrs in simulations of the convection zone only, resulted of 
the order of decades. 

\begin{figure}[htb]
\begin{center}
\includegraphics[width=0.78\textwidth]{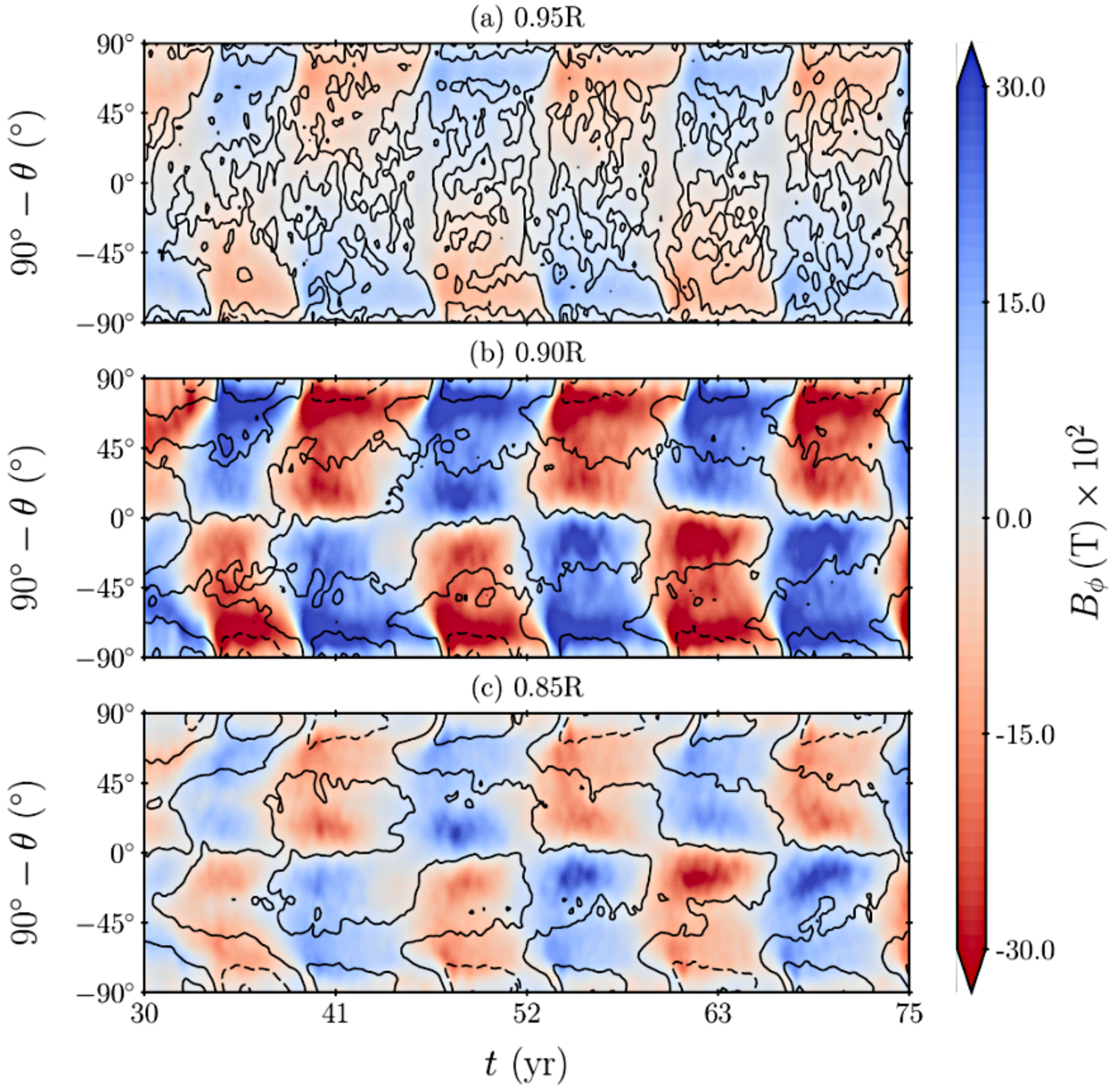}
\caption{Time-latitude butterfly diagram of one  simulation of the star
HD43587. The color contours show the toroidal mean magnetic field, $\mean{B}_{\phi}$.
The continuous (dashed) contour lines show the positive (negative) mean radial
magnetic field, $\mean{B}_{r}$.
\label{fig.mfb}}
\end{center}
\end{figure}

Figure~\ref{fig.mfb} shows the time-latitude butterfly diagram of a 
recent EULAG-MHD simulation of the solar analog HD43587 presented in 
\cite{ferreira+20}.
The color contours represent the strength of the
mean toroidal field, $\mean{B}_{\phi}$. The continuous (dashed) contour lines 
show the positive (negative) levels of the mean radial magnetic field, $\mean{B}_{r}$.
The panels (a), (b) and (c) correspond to different depths, as indicated. 
At $0.85 R_{*}$, where $R_{*} = 1.19\Rs$ is the radius of the star, the
butterfly diagram shows qualitative similarities with the solar one, i.e., 
there is strong magnetic field migrating equatorward at lower latitudes and
poleward migration at higher latitudes.  This configuration is likely
the result of magnetic field emerging from the tacholine at lower latitudes,
(see also the upper-left panel of Fig.~\ref{fig.mfb} which shows a snapshot
of the magnetic field configuration in the meridional, $\theta-r$, plane),
followed by poleward advection at the upper part of the domain. At $0.95 R_{*}$,
upper boundary of the simulation, only the polar branch is evident (this
stage corresponds to the lower-left panel of  Fig.~\ref{fig.mfb}).
\begin{figure}[htb]
\begin{center}
\includegraphics[width=0.4\textwidth]{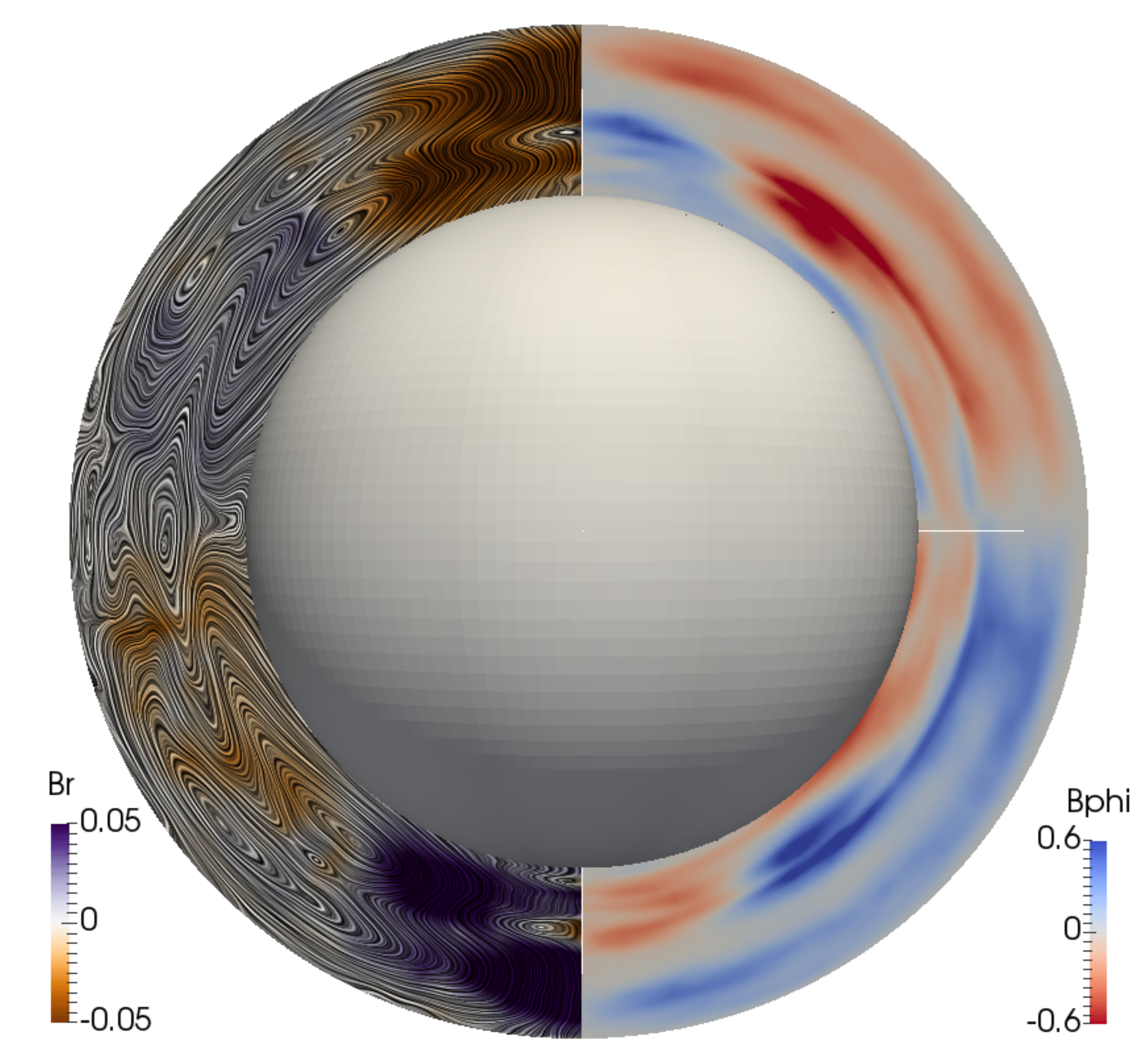}
\includegraphics[width=0.4\textwidth]{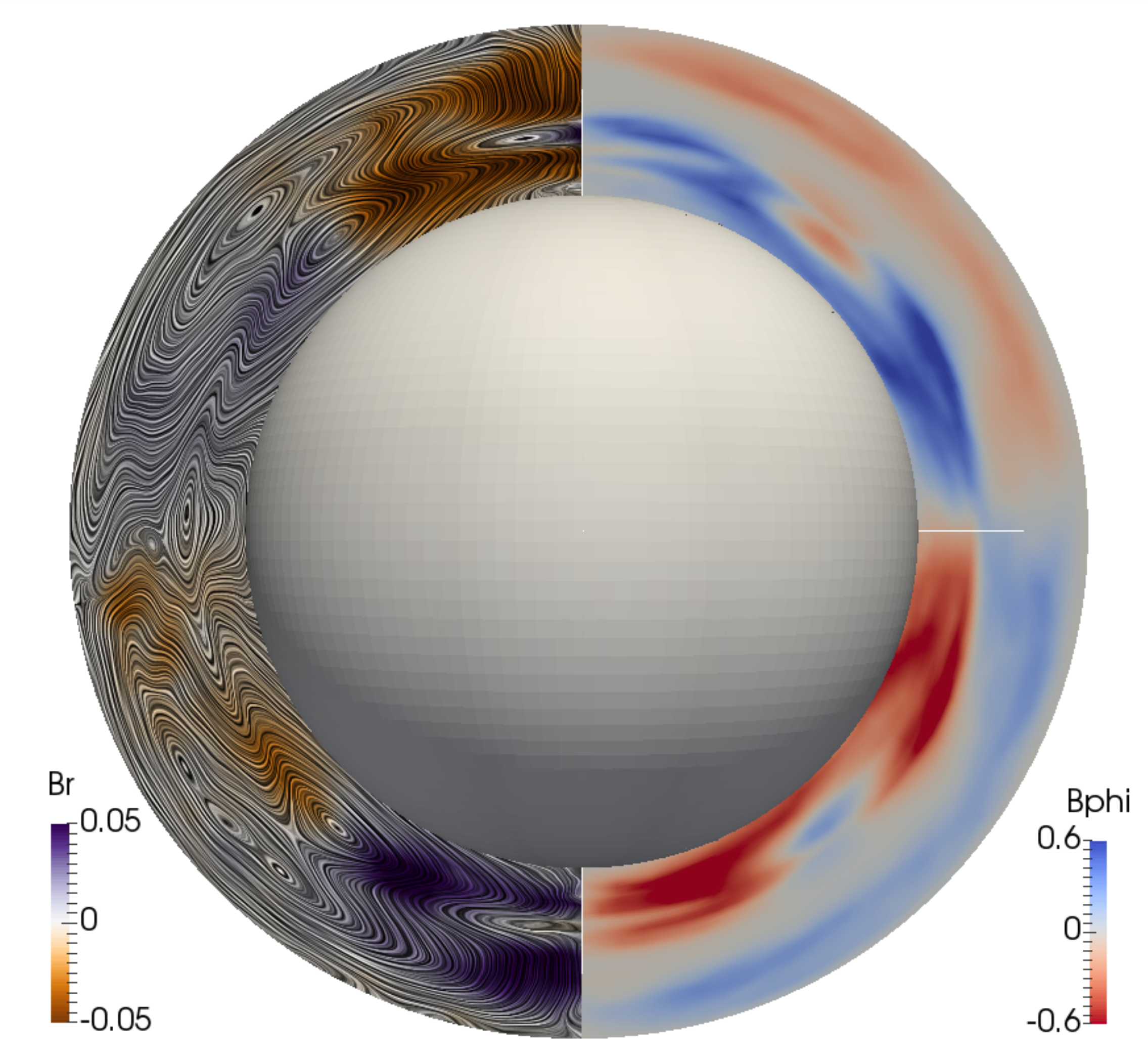}\\
\includegraphics[width=0.4\textwidth]{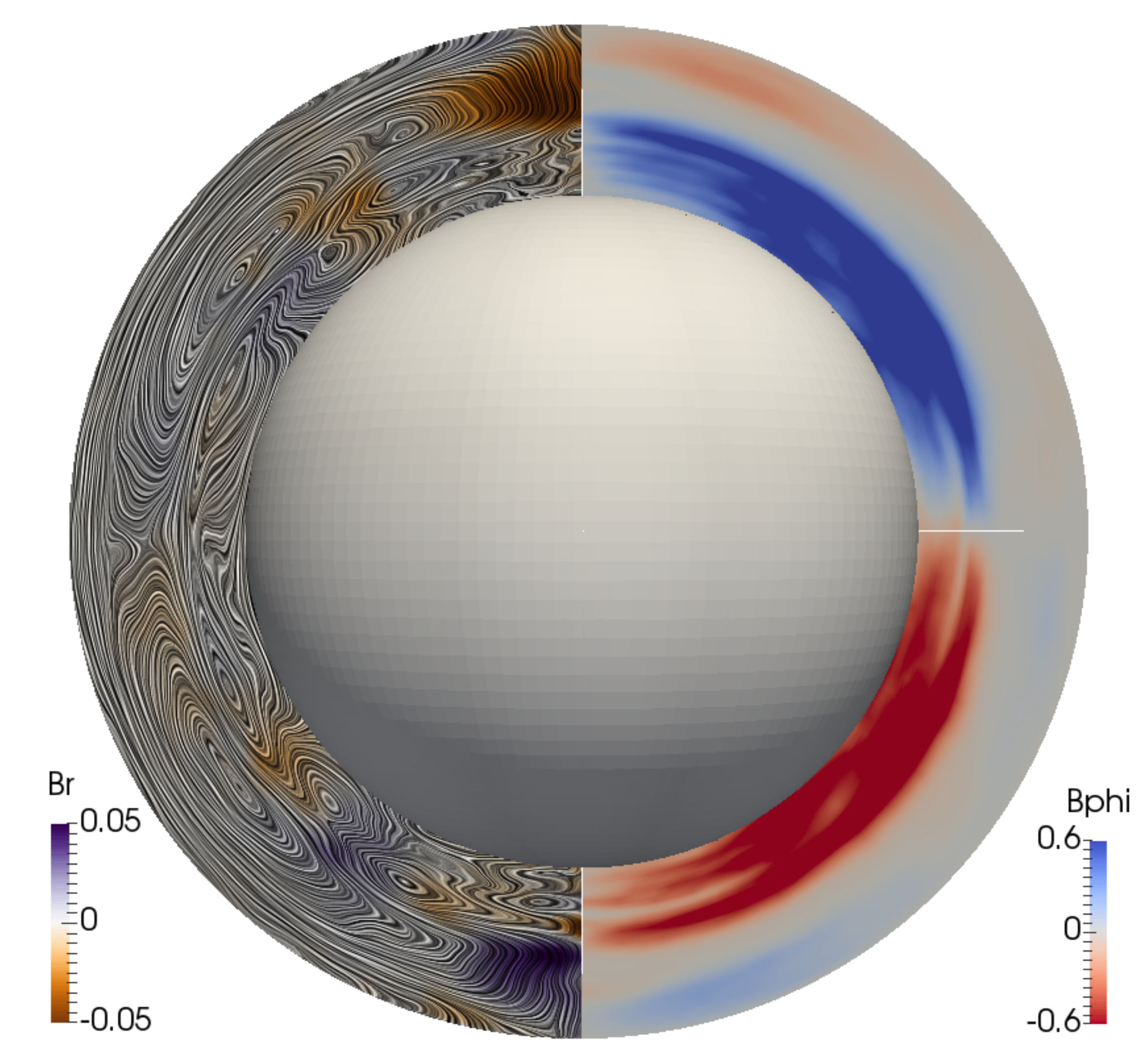}
\includegraphics[width=0.4\textwidth]{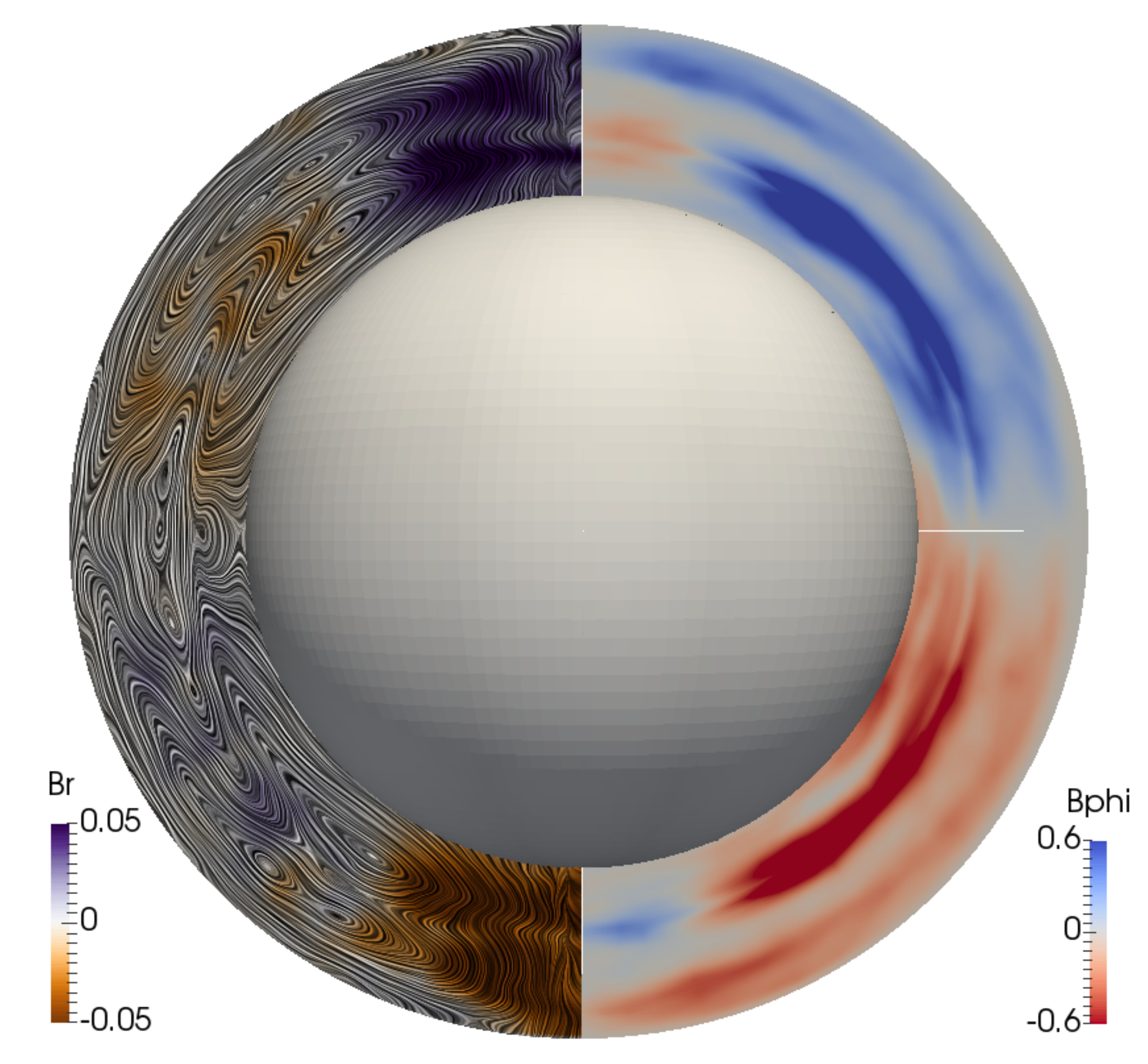}
\caption{Snapshots presenting half magnetic cycle in the EULAG-MHD simulation P25
presented in \cite{ferreira+20}.  On the left of each snapshot, the LIC representation 
depicts the distribution of the poloidal field lines with the color indicating the magnitude 
of the mean radial field, $\bar{B}_{r}(t,\theta,r)$.  The colored contours on the right
quadrants correspond to the azimuthal mean magnetic field, $\bar{B}_{\varphi}(t,\theta,r)$.  
\label{fig.mfs}}
\end{center}
\end{figure}

\subsection{Mean-field interpretation}
\label{s.mfd}

Similar to the theory of the $\Lambda$-effect, a mean-field model can be
used to describe the evolution of the large-scale magnetic field, $\mean{{\bm B}}$ 
in stellar interiors \citep{Pa55,SKR66}.  After separation of the turbulent and the 
large scales, the induction equation (Eq.~\ref{equ:in}) becomes,
\begin{equation}
\frac{\partial \mean{\bm{B}}}{\partial t} = \nabla \times (\mean{\bm{u}} \times \mean{\bm{B}})
                                          + \nabla \times (\alpha \mean{\bm{B}}) 
                                          - \nabla \times (\eta \nabla \times \mean{\bm{B}}),
\label{eq.mfie}
\end{equation}
where $\mean{\bm {B}}=(\mean{B}_r, \mean{B}_{\theta},\mean{B}_{\phi})$
and 
$\mean{\bm {u}} =  (\mean{u}_r, \mean{u}_{\theta},\mean{u}_{\phi})$ 
are the averaged magnetic and velocity fields. 
In the second term on the RHS, the $\alpha$ term
stems for the $\alpha$-effect which has kinetic and magnetic contributions,
namely $\alpha=\alpha_{\rm k}+\alpha_{\rm m}$.  These terms are the first
order correlations of the turbulent electromotive force, 
\begin{equation}
\mean{\cal {\bm E}} = \mean{{\bm u}' \times {\bm B}'} \simeq \alpha \mean{\bm B} - 
\etat \mu_0  \nabla \times \mean{\bm B} \;,
\label{equ:emf}
\end{equation}

\noindent where ${\bm u}'$ and
${\bm B}'$ are the turbulent velocity and magnetic fields and $\etat$ is
the turbulent magnetic diffusivity.  The $\alpha$ terms induce the large-scale magnetic 
field from small scale helical motions and currents. Under suitable closure models 
\citep[e.g.,][]{PFL76,Moffatt78}, the 
it may be expressed as:
\begin{equation}
\alpha = \alpha_{\rm k} + \alpha_{\rm m} = -\frac{\tau_c}{3}\brac{{\bm \omega'} \cdot {\bm u'}} 
                  + \frac{\tau_c}{3}\brac{{\bm j'} \cdot {\bm B'}}/\rho \;,
\label{equ:alpha}
\end{equation}
where ${\bm \omega}'= \nabla \times {\bm u}'$ and ${\bm j}'= \nabla \times {\bm B}'$ are
the small scale vorticity and current, respectively, and $\tau_c$ is the convective
turnover time of the turbulent motions.  

In the third term in Eq.~(\ref{eq.mfie}), $\eta = \eta_m + \etat$ is the
sum of the molecular and the turbulent magnetic diffusivities. The turbulent 
contribution is dominant by several orders of magnitude and under the first
order smoothed approximation (FOSA) may be computed as
\begin{equation}
\etat = \frac{1}{3} \tau_c u^{\prime 2} \;.
\label{eq.etat}
\end{equation}
 
Either the terms in the electromotive force (Eq.~\ref{equ:emf}) or those
in Eqs.~(\ref{equ:alpha} and \ref{eq.etat}) may be estimated
from the global simulations output allowing to search for a theoretical interpretation
of complex results. This has been done for the EULAG-MHD simulations of \cite{GCS10} by 
using a method for multidimensional regression \citep{RCGS11}. Their results indicate
that the process is consistent with an $\alpha^2\Omega$ dynamo \citep[see also][]{simard+13}.
For pencil-code dynamo results \cite{WKKB14} computed the dynamo coefficients,
$\alpha$ and $\Omega$, and proved that the evolution of the magnetic field follows the
\cite{Pa55}-\cite{Y75} sign rule. More recently \cite{Warnecke+18}
used the test-field method \citep{SRSRC05,BRS08} to perform a fiducial prediction
of the electromotive force coefficients.  For the profiles and amplitude of the 
coefficients, they conclude that the dominant dynamo in pencil-code simulations
is of $\alpha\Omega$ type with a localized $\alpha^2$ dynamo.

\begin{figure}[htb]
\begin{center}
\includegraphics[width=0.325\textwidth]{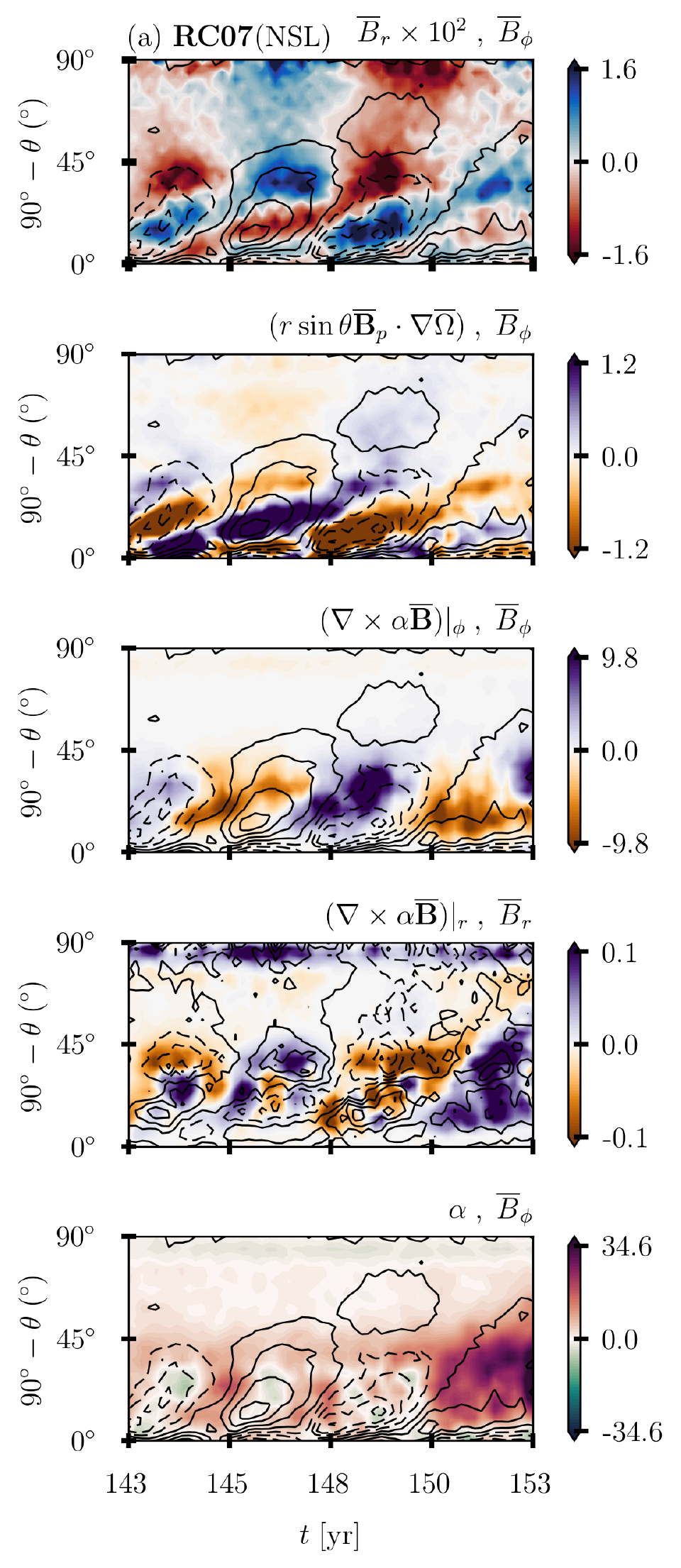}
\includegraphics[width=0.325\textwidth]{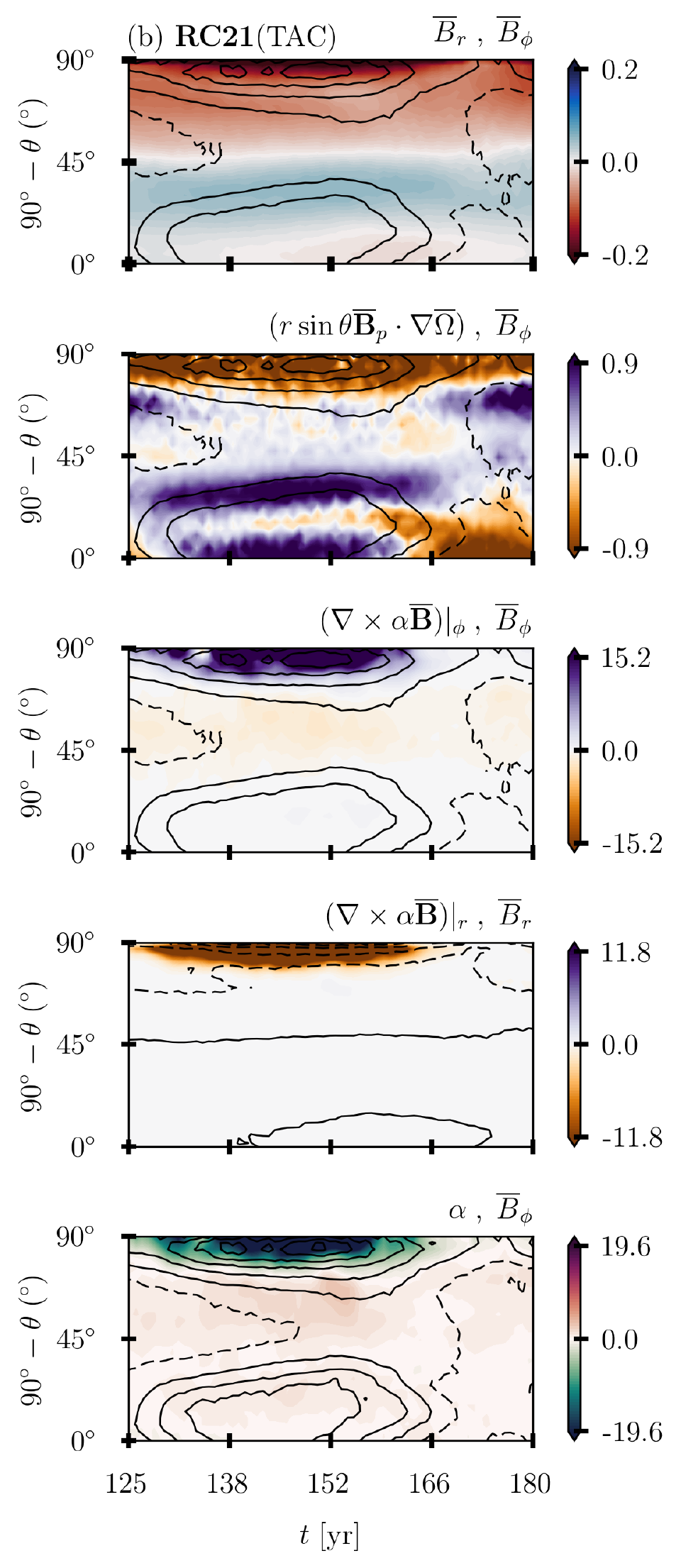}
\includegraphics[width=0.325\textwidth]{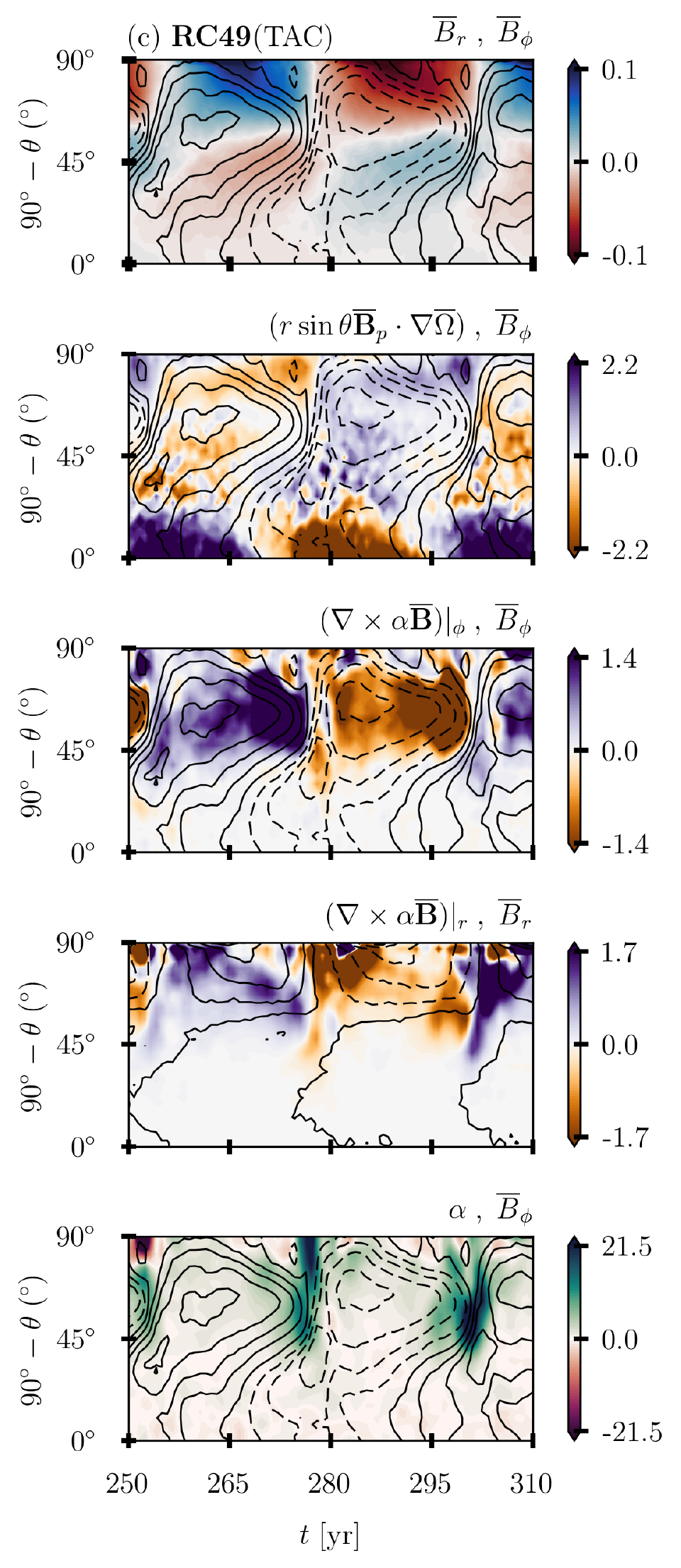}
\caption{Mean-field sources of large-scale  magnetic fields in some 
characteristic 
simulations presented in \cite{GZSDKM19}.  The columns correspond to
simulations with a cycle period of 7 (a), 21 (b) and 49 (c) days, respectively.  
All the quantities are averaged in longitude and over a radial extent in 
the near-surface layer (a) or in the tachocline (b and c).
From top to bottom the panels display the radial, $\bar{B}_r$ (colored contours) and toroidal,
$\bar{B}_{\phi}$ (contour lines) mean magnetic fields; the sources of the toroidal field, 
$r \sin \theta (\bar{B}_p \cdot \nabla) \bar{\Omega}$ and 
$(\nabla \times \alpha \bar{\bf B})|_{\phi}$ (colored contours), 
both compared with $\bar{B}_{\phi}$; 
the source of the radial field,  $(\nabla \times \alpha \bar{\bf B})|_{r}$ (color), 
compared with $\bar{B}_r$ (contour lines); and $\alpha = \alpha_k + \alpha_m$ 
(colored contours),  compared with $\bar{B}_{\phi}$. 
In the color map the dimensions are [T] for the magnetic field, 
$10^{-8}$ [T/s] for the source terms, and [m/s] for the $\alpha$-effect, respectively.  
Adapted from \cite{GZSDKM19}. 
\label{fig.src}}
\end{center}
\end{figure}

The FOSA coefficients of Eqs.~(\ref{equ:alpha} and \ref{eq.etat}) were computed by
\cite{GZSDKM19} for EULAG-MHD simulations including both, radiative and a convective
layer; having solar-like ambient state and rotating at different rates. The results
showed that for fast rotation the dominant dynamo is of $\alpha \Omega$ type
and operates at the convection zone
(Fig.~\ref{fig.src}(a)).
For intermediate and slow rotation rates, 
the dominant dynamo operates locally at the tachocline and is of $\alpha^2 \Omega$
type (Fig.~\ref{fig.src}(b and c)). The $\alpha$-effect contributes to the amplification 
of the toroidal field at intermediate to high latitudes while the rotational 
shear generates toroidal field at the equator. The results indicated that the appearance
of cycles in tachocline dynamos is related to the strength of the equatorial shear.
The most striking result observed in the simulations is that the $\alpha$-effect, at and 
below the tachocline, has magnetic origin, i.e., the inductive contribution of small-scale 
current helicity (second term in Eq.~\ref{equ:alpha}) is dominant over the induction of 
magnetic field by small-scale helical eddies.  

\begin{figure}[htb]
\begin{center}
\includegraphics[width=0.62\textwidth]{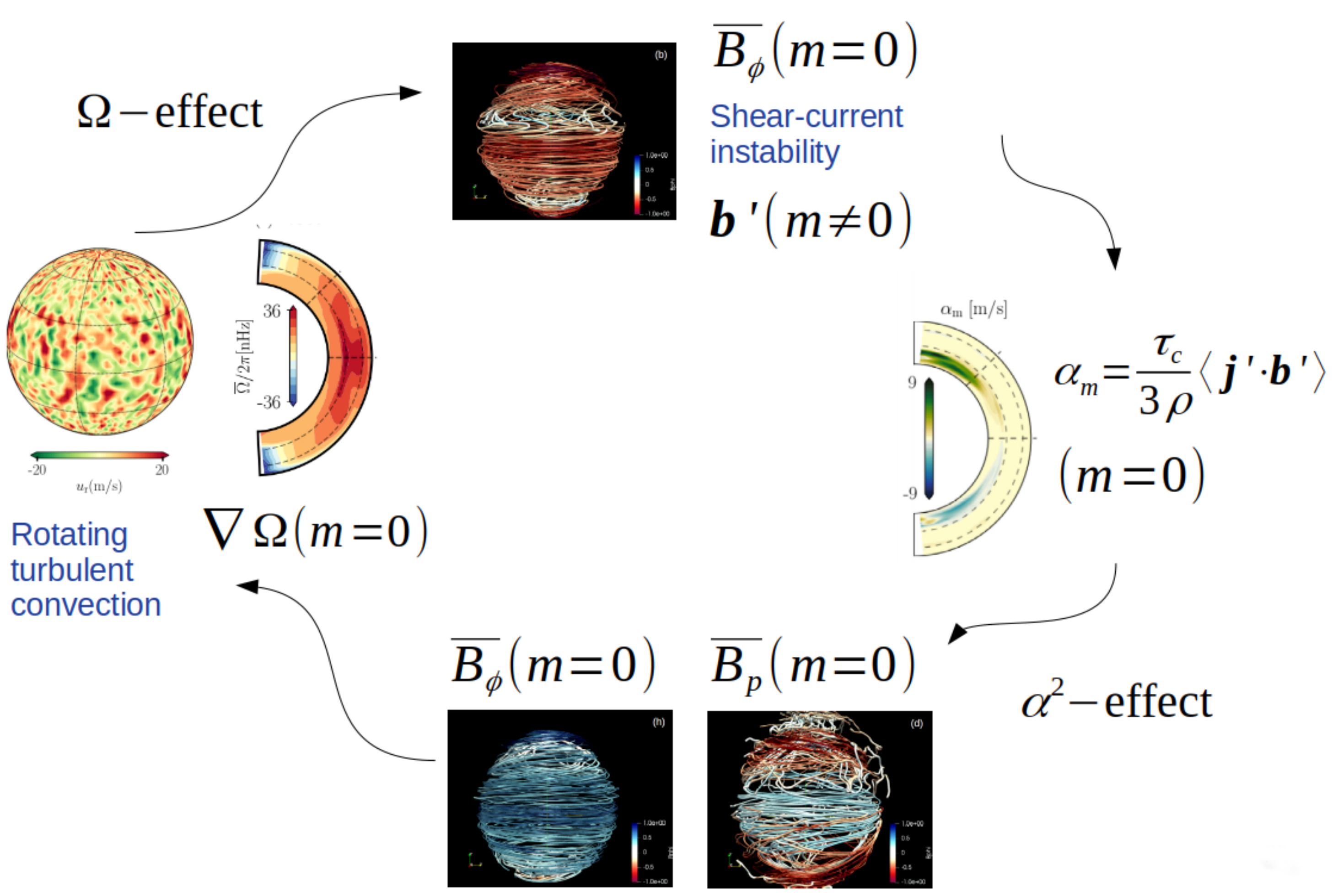}
\caption{Dynamo loop for oscillatory $\alpha^2\Omega$ dynamos operating
in the simulations of \cite{GZSDKM19}.
\label{fig.floop}}
\end{center}
\end{figure}

The magnetic loop observed in the deep seated $\alpha^2 \Omega$ dynamos
described above is illustrated in Fig.~\ref{fig.floop}. Turbulent convection is 
necessary to generate and sustain the differential rotation (left panel).  
It supplies the strong shear necessary for the $\Omega$-effect to 
develop a large-scale toroidal magnetic field, $\mean{B}_{\phi}$ (top).  
The generated layer of toroidal field is unstable to the shear 
present in the region and decays in non-axisymmetric modes  
($m \neq 0$),   whose collective contribution, in turn, develop an
axisymmetric magnetic $\alpha$-effect (right). This $\alpha$-effect
contributes to the generation of both, toroidal 
and poloidal mean magnetic fields.  Between maxima 
and minima the developed magnetic tension speeds-up and slows-down
the azimuthal motions creating a pattern of torsional oscillations. 
The mechanism generating this pattern is described in detail in \cite{GSDKM16b}. 

Several instabilities are simultaneously operating in the cycle depicted 
in Fig.~\ref{fig.floop}. Disentangle them is a rather difficult task, thus, 
simple simulations are necessary to understand the processes separately.  
The most intriguing 
mechanism in this dynamo is perhaps the generation of small scale helicities
in the stable layer because of instabilities of the magnetic field. 
We have claimed that it is consistent with a shear-current instability 
where a decaying
toroidal field produces helical currents that allow the growing of the 
poloidal field.  This kind of instabilities was first proposed by \citep{tayler} and
studied further by several authors \citep[e.g.,][]{pitts,bu12apj,bour13}. 
In the presence of shear the instability was explored in detail by 
\cite{Cally_03,Miesch07}. They reported that the shear-current instability 
for a broad band of toroidal magnetic field results in modes where the field lines 
open in the form of a clamshell. When the initial field has two bands of opposite
polarity, the axis of the bands tilts in latitude.
Similarly to the Tayler instability, in these cases the fastest 
growing mode is $m=1$. The clamshell instability is observed in the simulations 
presented in \cite[][see their Fig. 10]{GZSDKM19}.  A cyclic dynamo based 
on Tayler-like instabilities has been previously proposed by \cite{spruit02,ZB07,Bonanno13},
however, the global EULAG-MHD simulations of \cite{GZSDKM19} were the first 
ones to develop and sustain the dynamo. 

To fully understand how these instabilities are the source of an 
$\alpha$-effect, \cite{GZSDKM19} explored the stabilizing properties of 
gravity to the Tayler instability using the EULAG-MHD code.  The results showed
that for weak gravitational stratification the instability has radial dependence, while for
larger stratification the unstable modes  develop in horizontal layers and might
have oscillatory behavior. The
growth rate depends on the local value of the Brunt-Väisäla (BV) frequency
(in turn depending on the gravity force).  An extension to this work
is currently being carried out including rotation and shear (Monteiro
et al. 2020, {\it in preparation}). 

\subsection{Stellar cycles}

The study of magnetism in stars different from the Sun is necessary to get a 
broad  understanding of 
the physics in stellar interiors as well as the interaction between
stars and planetary systems.  It may also provide hints and  constraints to unveil 
the  details of the solar dynamo. From the Mount Wilson data several stars
were found to have activity cycles \citep{Baliunas+95, SB99} evident in 
the ${\rm Ca} ~_{\rm II} \; {\rm H}$ and ${\rm K}$ spectral lines. Further
studies correlated the magnetic cycle period to the rotational period
\citep{Noyes+84b,BST98,BV07,BMM17}. Two main branches, classified as active (A) 
or inactive (I) depending also in the strength of the magnetic field, 
have been canonically discussed in the literature \citep{BST98,BV07,BMM17}. 
Both of them show a positive linear correlation between $P_{\rm cyc}$ and 
$P_{\rm rot}$.  The same two branches appear parallel when $\log (P_{\rm rot}/P_{\rm cyc})$ 
is compared with $\log \brac{R'_{HK}}$, where $\brac{R'_{HK}}=F'_{HK}/F_{\rm bol}$
is the mean fractional ${\rm Ca} ~_{\rm II} \; {\rm H}$ and ${\rm K}$ flux
normalized with the stellar bolometric flux, $F_{\rm bol}$ \citep{BST98,BMM17}.
\cite{Olspert+18} re-visited the Mount Wilson data computing cycle periods with
sophisticated techniques for analysis of periodic and quasi-periodic time series. 
Their results confirmed the existence of the two branches, A and I. They found, however, 
that the trend of the active branch is not positive but broadly distributed
and connects with a third transitional branch of activity 
\citep{Lehtinen+16}.

Over the last years few attempts have been done to explore the relation between
magnetic cycles and rotation by using global convective dynamo simulations.  \cite{SBCBdN17}
performed ILES simulations with the EULAG-MHD code. They found a negative correlation 
between $P_{\rm cyc}$ and $P_{\rm rot}$
and argue that the cycle period is established by the back-reaction of the magnetic
field over the differential rotation \citep{strugarek+18}. \cite{warnecke18} reported
a similar relation in pencil-code wedge simulations. They attributed, however,
the magnetic cycles to $\alpha\Omega$ dynamos operating at the base of the convection 
zone.  The resulting cycle periods follow the linear mean-field dynamo theory 
\citep[see e.g.,][]{Stix76}:
\begin{equation}
P_{\rm cyc} \propto (C_{\alpha} C_{\Omega})^{-1/2}\;,
\label{eq.mfp}
\end{equation}

\noindent
where $C_{\alpha}$ and  $C_{\Omega}$ are non-dimensional quantities comparing
the inductive effects of the $\alpha$ and the $\Omega$ effects, respectively, 
with the magnetic turbulent diffusion.  \cite{warnecke18} argues that their 
simulations are reproducing the transitional branch. Another set of simulations
performed with the pencil-code cover the entire domain in the  $\varphi$ direction
but without reaching the poles \citep{Viviani+18}.  They explore fast
convective dynamos with rotations from $1$ to $30$ $\Omega_{\odot}$. Since
rotation breaks the convective cells, these simulations require high resolution. 
They found that the slow rotating cases fall in the I branch but having
a negative tilt.  On the other hand, the fast rotating cases result in 
non-axisymmetric dynamos with cycles falling in a different branch of 
activity representative of superactive stars.
\begin{figure}[htb]
\begin{center}
\includegraphics[width=0.72\textwidth]{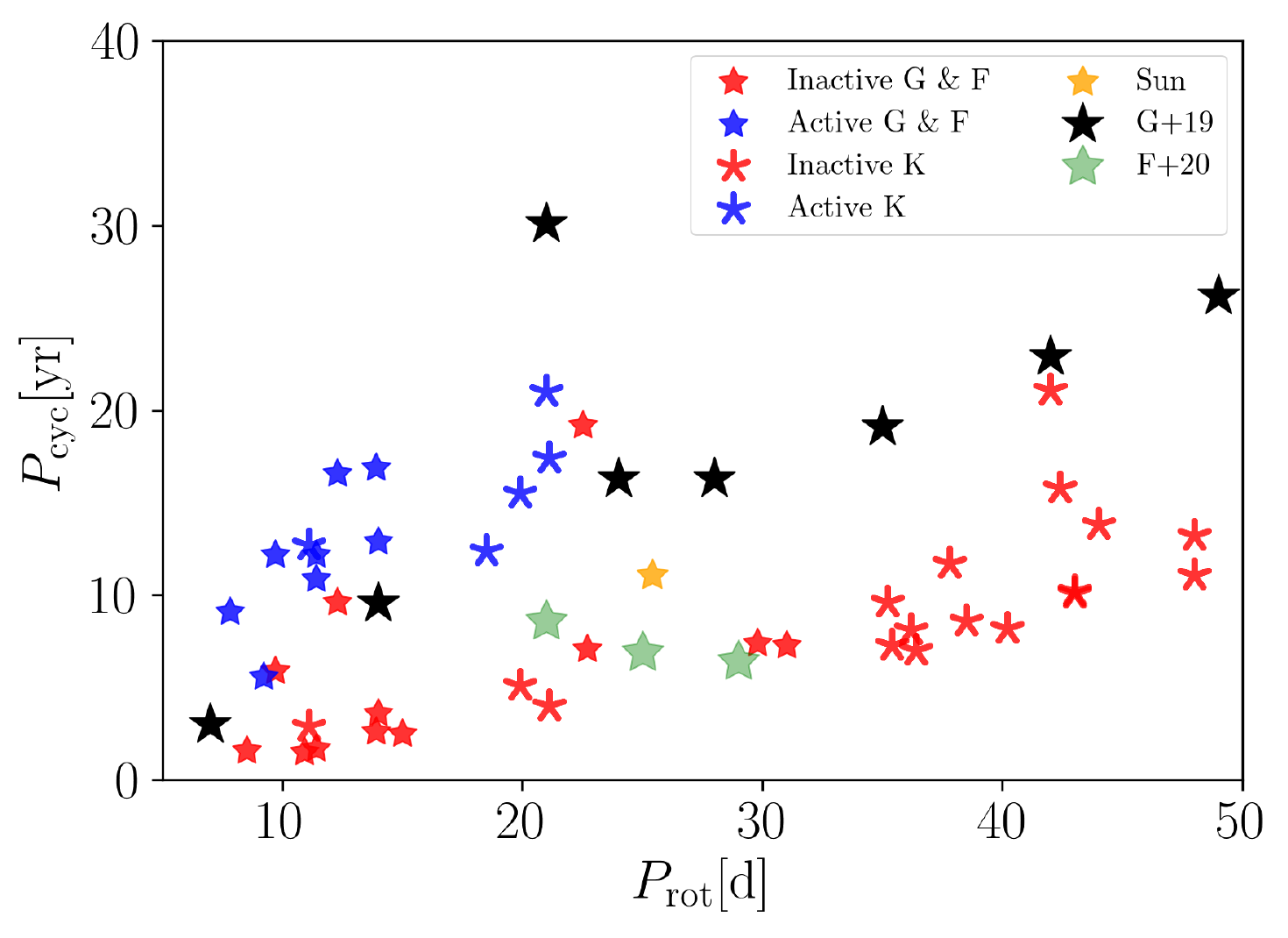}
\caption{Magnetic cycle period of rotation, $P_{\rm cyc}$  as a function of the rotational
period $P_{\rm rot}$.  The black stars correspond to the 
simulations presented in  \cite{GZSDKM19}, the green stars correspond to the 
simulations of the star HD43586 presented in \cite{ferreira+20}. 
\label{fig.prpc}}
\end{center}
\end{figure}

All these cyclic dynamo simulations have in common the absence of the strong
shear at the tachocline.  \cite{GSDKM16a}  found that long-lived strong magnetic fields
develop in models that include the tachocline and a stable layer underneath.  
A longer set of simulations was recently presented in \cite{GZSDKM19} with
dynamo solutions for rotational periods between 7 and 63 days. The results  
revealed two different types of dynamos:  $\alpha\Omega$  dynamos operating at the
bulk of the convection zone for fast rotating simulations, and depth seated
$\alpha^2\Omega$ dynamos for intermediate and slow rotating simulations. In between, 
there is a bifurcation point where deep seated oscillatory dynamo at  
lower latitudes coexists with a steady dynamo at the poles.  This point 
marks the level of equatorial shear necessary to generate tachocline cyclic dynamos.  
In both types of dynamos $P_{\rm cyc}$ increases with  $P_{\rm rot}$. The 
ones operating in the convection zone fall close to the active branch whereas
the tachocline dynamos fall in between the two branches (see black stars
in Fig.~\ref{fig.prpc}).  Although the deep seated magnetic cycles are 
consistent with $\alpha^2\Omega$ dynamos (as seen in \S\ref{s.mfd}), 
the cycle periods do not agree with the mean-field prediction (Eq.~\ref{eq.mfp}).  
The cycle is established once a balance between a non-linear dynamo and
tachocline instabilities is reached. A better understanding of this mechanism
is necessary to be able of predicting cycle lengths as a function
of stellar parameters. Furthermore, the numerical experiments presented in
\cite{ferreira+20} show that the simulations are 
rather sensitive to the profile o the ambient state, $\Theta_{\rm amb}$, 
described in section \S\ref{s.deu}.

In one hand, the level of superadiabaticity of the convection zone determines
the strength of the convection, and therefore the profile of differential rotation.
On the other hand, the level of subadiabaticity in the stable layer determines
the Brunt-Väisäla frequency and therefore the development of gravity waves in the 
stable layer.  These waves might be fundamental for the development of the 
instability of toroidal fields in radiative zones \citep{GdSBS19}.
We have implemented these ideas in the simulations of the solar analog 
HD43568 \citep{ferreira+20}.  The stratification profile in the
convection zone is set to fit the convective velocity obtained by the mixing length 
theory (MLT); in 
the stable layer it fits the BV frequency of the same structural model. In 
addition, we changed the magnetic boundary condition at the bottom of the
domain from radial field to perfect conductor. The results agree 
with the spectroscopic analysis indicating an activity cycle of 
$\sim 10$ years and low magnetic activity levels. Simulations for 21, 25 and
29 days results in cycles which fall in the inactive branch (see green stars in
Fig.~\ref{fig.prpc}). Further simulations with this improved ambient state
configuration are currently being performed for the Sun and other stars of the 
Mount Wilson sample. 

\section{Conclusions}
\label{s.c}

In this work we described the modeling of solar and stellar 
dynamos through self-consistent global simulations.  We argue that these are 
the most relevant tools to understand the development and sustain of stellar 
magnetic fields.  Special focus
is given to the recent ILES simulations performed with the anelastic 
solver EULAG-MHD.  A short discussion on the need of including 
SGS modeling in the current global modeling was also presented. 

In spite of the recent advances, the current models still are unable to 
reproduce the main characteristics of the solar dynamo nor the empirical
relations between magnetic cycle periods and magnetic field strength with
stellar parameters like the Rossby number. It is worth mentioning, however, 
that these relations still need to be revised with a large number 
of samples and longer observational missions. 

Regarding the development of mean flows, the result of global 
simulations with different numerical schemes show good agreement. 
Consequently, great advances have been done in the understanding of the buildup
of these flows. While the role of the Reynolds stresses is unanimously recognized 
as the responsible for the sustain of the differential rotation, 
the formation of meridional circulation has not yet a closed 
theoretical interpretation. It is attributed to the gyroscopic pumping
mechanism resulting from the thermal wind balance \citep{MH11}, or to 
deviations of this equilibrium given mostly by viscous forcing at boundary
layers \citep{kit13}. Most of global models tend to support the first 
alternative \citep[e.g.,][]{FM15,GSDKM16b}.

With respect to the development of mean magnetic fields, several questions 
still remain to be answered. Perhaps the most important are, first,  what global 
simulations have missed until now to reproduce solar-like patterns?
Second, what physical processes are occurring in the global simulations?
And third, how relevant are the tachoclines in the dynamo process?
About the first question, there is still room for exploration. For instance,
the role of the stratification, compressibility and a realistic upper 
magnetic boundary are yet to be studied.  In addition, simulations with
higher Reynolds numbers would likely be possible over the starting decade.
With respect to the second question, the work until now has proven that
the FOSA approximation might allow to distinguish what type of dynamo is occurring.
However, computing the full set of elements in the electromotive force tensors 
seems to be the appropriate alternative to
fully understand the dynamo process \citep[e.g.,][]{Warnecke+18}. 
The third question has been an issue of debate  over the last years. 
Doubts about the relevance of the layers of strong radial shear started
from the observational results of
\cite{Wright+16}. They reported that stars without tachoclines follow the
same scaling law between the magnetic field strength and $\Ro$ that stars 
with this layer.  On the other hand, recent observations of the
magnetic field topology in solar-type stars with different ages, show
a clear change at the age of the formation of the radiative
zone \citep{GDMHNHJ12}. The structural change corresponds to the shift 
from a mainly poloidal configuration to a magnetic field dominated by high
order modes. This is a clear indication of a change in the dynamo regime
at the age when the tachoclines appear.  
Whatever is the case, for the 
time being global models with and without tachoclines are able to develop
large-scale magnetic fields with polarity reversals. As these models are
better understood, further observational evidence will provide a clearest
scenario of stellar magnetism.  

\acknowledgments                                                                                
The author is thankful to the IAU and the Physics Department of UFMG for 
funding his participation in the symposium.  I'm grateful to all the 
collaborators who contributed to the publications discussed in this work. 
Special thanks to Rafaella Barbosa and Alexander Kosovichev for providing 
some of the figures. 
The simulations were performed in the NASA cluster Pleiades and Brazilian 
super computer SDumont of the National Laboratory of Scientific Computation 
(LNCC).

\bibliographystyle{aa}
\bibliography{bib}

\begin{thebibliography}{101}
\expandafter\ifx\csname natexlab\endcsname\relax\def\natexlab#1{#1}\fi

\bibitem[{{Antia} \& {Basu}(2001)}]{AB01}
{Antia}, H.~M. \& {Basu}, S. 2001, \apjl, 559, L67

\bibitem[{{Augustson} {et~al.}(2015){Augustson}, {Brun}, {Miesch}, \&
  {Toomre}}]{ABMT15}
{Augustson}, K., {Brun}, A.~S., {Miesch}, M., \& {Toomre}, J. 2015, \apj, 809,
  149

\bibitem[{{Baliunas} {et~al.}(1995){Baliunas}, {Donahue}, {Soon}, {Horne},
  {Frazer}, {Woodard-Eklund}, {Bradford}, {Rao}, {Wilson}, {Zhang}, {Bennett},
  {Briggs}, {Carroll}, {Duncan}, {Figueroa}, {Lanning}, {Misch}, {Mueller},
  {Noyes}, {Poppe}, {Porter}, {Robinson}, {Russell}, {Shelton}, {Soyumer},
  {Vaughan}, \& {Whitney}}]{Baliunas+95}
{Baliunas}, S.~L., {Donahue}, R.~A., {Soon}, W.~H., {et~al.} 1995, \apj, 438,
  269

\bibitem[{{Benomar} {et~al.}(2018){Benomar}, {Bazot}, {Nielsen}, {Gizon},
  {Sekii}, {Takata}, {Hotta}, {Hanasoge}, {Sreenivasan}, \&
  {Christensen-Dalsgaard}}]{Benomar+18}
{Benomar}, O., {Bazot}, M., {Nielsen}, M.~B., {et~al.} 2018, Science, 361, 1231

\bibitem[{{B{\"o}hm-Vitense}(2007)}]{BV07}
{B{\"o}hm-Vitense}, E. 2007, \apj, 657, 486

\bibitem[{{Bonanno}(2013)}]{Bonanno13}
{Bonanno}, A. 2013, \solphys, 287, 185

\bibitem[{{Bonanno} \& {Urpin}(2012)}]{bu12apj}
{Bonanno}, A. \& {Urpin}, V. 2012, \apj, 747, 137

\bibitem[{{Bonanno} \& {Urpin}(2013)}]{bour13}
{Bonanno}, A. \& {Urpin}, V. 2013, \apj, 766, 52

\bibitem[{{B{\"o}ning} {et~al.}(2017){B{\"o}ning}, {Roth}, {Jackiewicz}, \&
  {Kholikov}}]{boning+17}
{B{\"o}ning}, V. G.~A., {Roth}, M., {Jackiewicz}, J., \& {Kholikov}, S. 2017,
  \apj, 845, 2

\bibitem[{{Brandenburg} {et~al.}(2017){Brandenburg}, {Mathur}, \&
  {Metcalfe}}]{BMM17}
{Brandenburg}, A., {Mathur}, S., \& {Metcalfe}, T.~S. 2017, \apj, 845, 79

\bibitem[{{Brandenburg} {et~al.}(2008){Brandenburg}, {R{\"a}dler}, \&
  {Schrinner}}]{BRS08}
{Brandenburg}, A., {R{\"a}dler}, K.~H., \& {Schrinner}, M. 2008, \aap, 482, 739

\bibitem[{{Brandenburg} {et~al.}(1998){Brandenburg}, {Saar}, \&
  {Turpin}}]{BST98}
{Brandenburg}, A., {Saar}, S.~H., \& {Turpin}, C.~R. 1998, \apjl, 498, L51

\bibitem[{{Brown} {et~al.}(2010){Brown}, {Browning}, {Brun}, {Miesch}, \&
  {Toomre}}]{BBBMT10}
{Brown}, B.~P., {Browning}, M.~K., {Brun}, A.~S., {Miesch}, M.~S., \& {Toomre},
  J. 2010, \apj, 711, 424

\bibitem[{{Browning} {et~al.}(2006){Browning}, {Miesch}, {Brun}, \&
  {Toomre}}]{BMBT06}
{Browning}, M.~K., {Miesch}, M.~S., {Brun}, A.~S., \& {Toomre}, J. 2006, \apjl,
  648, L157

\bibitem[{{Brun} {et~al.}(2004){Brun}, {Miesch}, \& {Toomre}}]{BMT04}
{Brun}, A.~S., {Miesch}, M.~S., \& {Toomre}, J. 2004, \apj, 614, 1073

\bibitem[{{Brun} {et~al.}(2011){Brun}, {Miesch}, \& {Toomre}}]{BMT11}
{Brun}, A.~S., {Miesch}, M.~S., \& {Toomre}, J. 2011, \apj, 742, 79

\bibitem[{{Brun} {et~al.}(2017){Brun}, {Strugarek}, {Varela}, {Matt},
  {Augustson}, {Emeriau}, {DoCao}, {Brown}, \& {Toomre}}]{brun+17}
{Brun}, A.~S., {Strugarek}, A., {Varela}, J., {et~al.} 2017, \apj, 836, 192

\bibitem[{{Brun} \& {Toomre}(2002)}]{BT02}
{Brun}, A.~S. \& {Toomre}, J. 2002, \apj, 570, 865

\bibitem[{{Cally}(2003)}]{Cally_03}
{Cally}, P.~S. 2003, \mnras, 339, 957

\bibitem[{{Cossette} {et~al.}(2017){Cossette}, {Charbonneau}, {Smolarkiewicz},
  \& {Rast}}]{Cossette+17}
{Cossette}, J.-F., {Charbonneau}, P., {Smolarkiewicz}, P.~K., \& {Rast}, M.~P.
  2017, \apj, 841, 65

\bibitem[{{Elliott} \& {Smolarkiewicz}(2002)}]{ES02}
{Elliott}, J.~R. \& {Smolarkiewicz}, P.~K. 2002, International Journal for
  Numerical Methods in Fluids, 39, 855

\bibitem[{{Fan} \& {Fang}(2014)}]{FF14}
{Fan}, Y. \& {Fang}, F. 2014, \apj, 789, 35

\bibitem[{{Featherstone} \& {Miesch}(2015)}]{FM15}
{Featherstone}, N.~A. \& {Miesch}, M.~S. 2015, \apj, 804, 67

\bibitem[{{Ferreira} {et~al.}(2020){Ferreira}, {Barbosa}, {Castro}, {Guerrero},
  {de Almeida}, {Boumier}, \& {do Nascimento}}]{ferreira+20}
{Ferreira}, R.~R., {Barbosa}, R., {Castro}, M., {et~al.} 2020, Submited to
  A\&A.

\bibitem[{{Gastine} {et~al.}(2014){Gastine}, {Yadav}, {Morin}, {Reiners}, \&
  {Wicht}}]{GYMRW14}
{Gastine}, T., {Yadav}, R.~K., {Morin}, J., {Reiners}, A., \& {Wicht}, J. 2014,
  \mnras, 438, L76

\bibitem[{{Germano} {et~al.}(1991){Germano}, {Piomelli}, {Moin}, \&
  {Cabot}}]{GPMC91}
{Germano}, M., {Piomelli}, U., {Moin}, P., \& {Cabot}, W.~H. 1991, Physics of
  Fluids, 3, 1760

\bibitem[{{Ghizaru} {et~al.}(2010){Ghizaru}, {Charbonneau}, \&
  {Smolarkiewicz}}]{GCS10}
{Ghizaru}, M., {Charbonneau}, P., \& {Smolarkiewicz}, P.~K. 2010, ApJL, 715,
  L133

\bibitem[{{Gilman}(1977)}]{gilman77}
{Gilman}, P.~A. 1977, Geophysical and Astrophysical Fluid Dynamics, 8, 93

\bibitem[{{Gilman}(1983)}]{gilman83}
{Gilman}, P.~A. 1983, \apjs, 53, 243

\bibitem[{{Gilman} \& {Miller}(1981)}]{GM81}
{Gilman}, P.~A. \& {Miller}, J. 1981, \apjs, 46, 211

\bibitem[{{Glatzmaier}(1984)}]{glatz84}
{Glatzmaier}, G.~A. 1984, Journal of Computational Physics, 55, 461

\bibitem[{{Glatzmaier}(1985{\natexlab{a}})}]{glatz85a}
{Glatzmaier}, G.~A. 1985{\natexlab{a}}, \apj, 291, 300

\bibitem[{{Glatzmaier}(1985{\natexlab{b}})}]{glatz85b}
{Glatzmaier}, G.~A. 1985{\natexlab{b}}, Geophysical and Astrophysical Fluid
  Dynamics, 31, 137

\bibitem[{{Gregory} {et~al.}(2012){Gregory}, {Donati}, {Morin}, {Hussain},
  {Mayne}, {Hillenbrand}, \& {Jardine}}]{GDMHNHJ12}
{Gregory}, S.~G., {Donati}, J.-F., {Morin}, J., {et~al.} 2012, \apj, 755, 97

\bibitem[{Grinstein {et~al.}(2007)Grinstein, Margolin, \& Rider}]{Grinstein+07}
Grinstein, F., Margolin, L., \& Rider, W. 2007, Implicit Large Eddy Simulation:
  Computing Turbulent Fluid Dynamics (Cambridge University Press)

\bibitem[{{Guerrero} {et~al.}(2019{\natexlab{a}}){Guerrero}, {Del Sordo},
  {Bonanno}, \& {Smolarkiewicz}}]{GdSBS19}
{Guerrero}, G., {Del Sordo}, F., {Bonanno}, A., \& {Smolarkiewicz}, P.~K.
  2019{\natexlab{a}}, \mnras, 490, 4281

\bibitem[{{Guerrero} {et~al.}(2016{\natexlab{a}}){Guerrero}, {Smolarkiewicz},
  {de Gouveia Dal Pino}, {Kosovichev}, \& {Mansour}}]{GSDKM16a}
{Guerrero}, G., {Smolarkiewicz}, P.~K., {de Gouveia Dal Pino}, E.~M.,
  {Kosovichev}, A.~G., \& {Mansour}, N.~N. 2016{\natexlab{a}}, \apj, 819, 104

\bibitem[{{Guerrero} {et~al.}(2016{\natexlab{b}}){Guerrero}, {Smolarkiewicz},
  {de Gouveia Dal Pino}, {Kosovichev}, \& {Mansour}}]{GSDKM16b}
{Guerrero}, G., {Smolarkiewicz}, P.~K., {de Gouveia Dal Pino}, E.~M.,
  {Kosovichev}, A.~G., \& {Mansour}, N.~N. 2016{\natexlab{b}}, \apjl, 828, L3

\bibitem[{{Guerrero} {et~al.}(2013){Guerrero}, {Smolarkiewicz}, {Kosovichev},
  \& {Mansour}}]{GSKM13b}
{Guerrero}, G., {Smolarkiewicz}, P.~K., {Kosovichev}, A.~G., \& {Mansour},
  N.~N. 2013, \apj, 779, 176

\bibitem[{{Guerrero} {et~al.}(2019{\natexlab{b}}){Guerrero}, {Zaire},
  {Smolarkiewicz}, {de Gouveia Dal Pino}, {Kosovichev}, \&
  {Mansour}}]{GZSDKM19}
{Guerrero}, G., {Zaire}, B., {Smolarkiewicz}, P.~K., {et~al.}
  2019{\natexlab{b}}, \apj, 880, 6

\bibitem[{{Haugen} {et~al.}(2004){Haugen}, {Brandenburg}, \& {Dobler}}]{HB04}
{Haugen}, N.~E., {Brandenburg}, A., \& {Dobler}, W. 2004, \pre, 70, 016308

\bibitem[{{Haugen} \& {Brandenburg}(2006)}]{HB06}
{Haugen}, N.~E.~L. \& {Brandenburg}, A. 2006, Physics of Fluids, 18, 075106

\bibitem[{{Held} \& {Suarez}(1994)}]{HS94}
{Held}, I.~M. \& {Suarez}, M.~J. 1994, Bulletin of the American Meteorological
  Society, 75, 1825

\bibitem[{{Hotta} {et~al.}(2016){Hotta}, {Rempel}, \& {Yokoyama}}]{Hotta+16}
{Hotta}, H., {Rempel}, M., \& {Yokoyama}, T. 2016, Science, 351, 1427

\bibitem[{{Howard} \& {Labonte}(1980)}]{HL80}
{Howard}, R. \& {Labonte}, B.~J. 1980, \apjl, 239, L33

\bibitem[{{Kaneda} {et~al.}(2003){Kaneda}, {Ishihara}, {Yokokawa}, {Itakura},
  \& {Uno}}]{Kaneda03}
{Kaneda}, Y., {Ishihara}, T., {Yokokawa}, M., {Itakura}, K., \& {Uno}, A. 2003,
  Physics of Fluids, 15, L21

\bibitem[{{K{\"a}pyl{\"a}} {et~al.}(2010){K{\"a}pyl{\"a}}, {Korpi},
  {Brandenburg}, {Mitra}, \& {Tavakol}}]{KKBMT10}
{K{\"a}pyl{\"a}}, P.~J., {Korpi}, M.~J., {Brandenburg}, A., {Mitra}, D., \&
  {Tavakol}, R. 2010, Astronomische Nachrichten, 331, 73

\bibitem[{{K{\"a}pyl{\"a}} {et~al.}(2012){K{\"a}pyl{\"a}}, {Mantere}, \&
  {Brandenburg}}]{KMB12}
{K{\"a}pyl{\"a}}, P.~J., {Mantere}, M.~J., \& {Brandenburg}, A. 2012, ApJL,
  755, L22

\bibitem[{{K{\"a}pyl{\"a}} {et~al.}(2011){K{\"a}pyl{\"a}}, {Mantere},
  {Guerrero}, {Brandenburg}, \& {Chatterjee}}]{KMGBC11}
{K{\"a}pyl{\"a}}, P.~J., {Mantere}, M.~J., {Guerrero}, G., {Brandenburg}, A.,
  \& {Chatterjee}, P. 2011, \aap, 531, A162

\bibitem[{{Karak} {et~al.}(2015){Karak}, {K{\"a}pyl{\"a}}, {K{\"a}pyl{\"a}},
  {Brandenburg}, {Olspert}, \& {Pelt}}]{KKBOP15}
{Karak}, B.~B., {K{\"a}pyl{\"a}}, P.~J., {K{\"a}pyl{\"a}}, M.~J., {et~al.}
  2015, \aap, 576, A26

\bibitem[{{Karak} {et~al.}(2014){Karak}, {Kitchatinov}, \& {Choudhuri}}]{KKC14}
{Karak}, B.~B., {Kitchatinov}, L.~L., \& {Choudhuri}, A.~R. 2014, \apj, 791, 59

\bibitem[{{K{\H{o}}v{\'a}ri} {et~al.}(2017){K{\H{o}}v{\'a}ri}, {Ol{\'a}h},
  {Kriskovics}, {Vida}, {Forg{\'a}cs-Dajka}, \& {Strassmeier}}]{kovari+17}
{K{\H{o}}v{\'a}ri}, Z., {Ol{\'a}h}, K., {Kriskovics}, L., {et~al.} 2017,
  Astronomische Nachrichten, 338, 903

\bibitem[{{Kitchatinov}(2013)}]{kit13}
{Kitchatinov}, L.~L. 2013, in IAU Symposium, Vol. 294, Solar and Astrophysical
  Dynamos and Magnetic Activity, ed. A.~G. {Kosovichev}, E.~{de Gouveia Dal
  Pino}, \& Y.~{Yan}, 399--410

\bibitem[{{Kosovichev} \& {Schou}(1997)}]{KS97}
{Kosovichev}, A.~G. \& {Schou}, J. 1997, \apjl, 482, L207

\bibitem[{{Lehtinen} {et~al.}(2016){Lehtinen}, {Jetsu}, {Hackman}, {Kajatkari},
  \& {Henry}}]{Lehtinen+16}
{Lehtinen}, J., {Jetsu}, L., {Hackman}, T., {Kajatkari}, P., \& {Henry}, G.~W.
  2016, \aap, 588, A38

\bibitem[{{Liang} {et~al.}(2018){Liang}, {Gizon}, {Birch}, {Duvall}, \&
  {Rajaguru}}]{liang+18}
{Liang}, Z.-C., {Gizon}, L., {Birch}, A.~C., {Duvall}, T.~L., \& {Rajaguru},
  S.~P. 2018, \aap, 619, A99

\bibitem[{Margolin(2019)}]{Margolin2019}
Margolin, L.~G. 2019, Shock Waves, 29, 27

\bibitem[{{Margolin} \& {Rider}(2002)}]{MR02}
{Margolin}, L.~G. \& {Rider}, W.~J. 2002, International Journal for Numerical
  Methods in Fluids, 39, 821

\bibitem[{{Masada} {et~al.}(2013){Masada}, {Yamada}, \& {Kageyama}}]{MYK13}
{Masada}, Y., {Yamada}, K., \& {Kageyama}, A. 2013, \apj, 778, 11

\bibitem[{{Matilsky} {et~al.}(2019){Matilsky}, {Hindman}, \& {Toomre}}]{MHBT19}
{Matilsky}, L.~I., {Hindman}, B.~W., \& {Toomre}, J. 2019, \apj, 871, 217

\bibitem[{{Miesch} {et~al.}(2006){Miesch}, {Brun}, \& {Toomre}}]{MBT06}
{Miesch}, M.~S., {Brun}, A.~S., \& {Toomre}, J. 2006, \apj, 641, 618

\bibitem[{{Miesch} {et~al.}(2007){Miesch}, {Gilman}, \& {Dikpati}}]{Miesch07}
{Miesch}, M.~S., {Gilman}, P.~A., \& {Dikpati}, M. 2007, \apjs, 168, 337

\bibitem[{{Miesch} \& {Hindman}(2011)}]{MH11}
{Miesch}, M.~S. \& {Hindman}, B.~W. 2011, \apj, 743, 79

\bibitem[{{Moffatt}(1978)}]{Moffatt78}
{Moffatt}, H.~K. 1978, {Magnetic field generation in electrically conducting
  fluids}

\bibitem[{{Noyes} {et~al.}(1984){Noyes}, {Weiss}, \& {Vaughan}}]{Noyes+84b}
{Noyes}, R.~W., {Weiss}, N.~O., \& {Vaughan}, A.~H. 1984, \apj, 287, 769

\bibitem[{{Olspert} {et~al.}(2018){Olspert}, {Lehtinen}, {K{\"a}pyl{\"a}},
  {Pelt}, \& {Grigorievskiy}}]{Olspert+18}
{Olspert}, N., {Lehtinen}, J.~J., {K{\"a}pyl{\"a}}, M.~J., {Pelt}, J., \&
  {Grigorievskiy}, A. 2018, \aap, 619, A6

\bibitem[{{Parker}(1955)}]{Pa55}
{Parker}, E.~N. 1955, \apj, 122, 293

\bibitem[{{Passos} {et~al.}(2017){Passos}, {Miesch}, {Guerrero}, \&
  {Charbonneau}}]{PMCG16}
{Passos}, D., {Miesch}, M., {Guerrero}, G., \& {Charbonneau}, P. 2017, \aap,
  607, A120

\bibitem[{{Pipin} \& {Kosovichev}(2018)}]{PK18}
{Pipin}, V.~V. \& {Kosovichev}, A.~G. 2018, \apj, 854, 67

\bibitem[{{Pitts} \& {Tayler}(1985)}]{pitts}
{Pitts}, E. \& {Tayler}, R.~J. 1985, \mnras, 216, 139

\bibitem[{{Pouquet} {et~al.}(1976){Pouquet}, {Frisch}, \& {Leorat}}]{PFL76}
{Pouquet}, A., {Frisch}, U., \& {Leorat}, J. 1976, Journal of Fluid Mechanics,
  77, 321

\bibitem[{{Prusa} {et~al.}(2008){Prusa}, {Smolarkiewicz}, \&
  {Wyszogrodzki}}]{PSW08}
{Prusa}, J.~M., {Smolarkiewicz}, P.~K., \& {Wyszogrodzki}, A.~A. 2008, Comput.
  Fluids, 37, 1193

\bibitem[{{Racine} {et~al.}(2011){Racine}, {Charbonneau}, {Ghizaru}, {Bouchat},
  \& {Smolarkiewicz}}]{RCGS11}
{Racine}, {\'E}., {Charbonneau}, P., {Ghizaru}, M., {Bouchat}, A., \&
  {Smolarkiewicz}, P.~K. 2011, \apj, 735, 46

\bibitem[{{Reinhold} \& {Gizon}(2015)}]{RG15}
{Reinhold}, T. \& {Gizon}, L. 2015, \aap, 583, A65

\bibitem[{{Ruediger}(1989)}]{rudiger89}
{Ruediger}, G. 1989, {Differential rotation and stellar convection. Sun and the
  solar stars}

\bibitem[{{Saar} \& {Brandenburg}(1999)}]{SB99}
{Saar}, S.~H. \& {Brandenburg}, A. 1999, \apj, 524, 295

\bibitem[{{Schad} {et~al.}(2013){Schad}, {Timmer}, \& {Roth}}]{str13}
{Schad}, A., {Timmer}, J., \& {Roth}, M. 2013, \apjl, 778, L38

\bibitem[{{Schou} {et~al.}(1998){Schou}, {Antia}, {Basu}, {Bogart}, {Bush},
  {Chitre}, {Christensen-Dalsgaard}, {Di Mauro}, {Dziembowski}, {Eff-Darwich},
  {Gough}, {Haber}, {Hoeksema}, {Howe}, {Korzennik}, {Kosovichev}, {Larsen},
  {Pijpers}, {Scherrer}, {Sekii}, {Tarbell}, {Title}, {Thompson}, \&
  {Toomre}}]{SCHOU98}
{Schou}, J., {Antia}, H.~M., {Basu}, S., {et~al.} 1998, \apj, 505, 390

\bibitem[{{Schrinner} {et~al.}(2005){Schrinner}, {R{\"a}dler}, {Schmitt},
  {Rheinhardt}, \& {Christensen}}]{SRSRC05}
{Schrinner}, M., {R{\"a}dler}, K.-H., {Schmitt}, D., {Rheinhardt}, M., \&
  {Christensen}, U. 2005, Astronomische Nachrichten, 326, 245

\bibitem[{{Simard} {et~al.}(2013){Simard}, {Charbonneau}, \&
  {Bouchat}}]{simard+13}
{Simard}, C., {Charbonneau}, P., \& {Bouchat}, A. 2013, \apj, 768, 16

\bibitem[{{Smagorinsky}(1963)}]{Sma63}
{Smagorinsky}, J. 1963, Monthly Weather Review, 91, 99

\bibitem[{{Smolarkiewicz}(2006)}]{S06}
{Smolarkiewicz}, P.~K. 2006, International Journal for Numerical Methods in
  Fluids, 50, 1123

\bibitem[{Smolarkiewicz \& Charbonneau(2013)}]{SC13}
Smolarkiewicz, P.~K. \& Charbonneau, P. 2013, J. Comput. Phys., 236, 608

\bibitem[{{Spitzer}(1962)}]{spitzer62}
{Spitzer}, L. 1962, {Physics of Fully Ionized Gases}

\bibitem[{{Spruit}(2002)}]{spruit02}
{Spruit}, H.~C. 2002, \aap, 381, 923

\bibitem[{{Steenbeck} {et~al.}(1966){Steenbeck}, {Krause}, \&
  {R{\"a}dler}}]{SKR66}
{Steenbeck}, M., {Krause}, F., \& {R{\"a}dler}, K.-H. 1966, Zeitschrift
  Naturforschung Teil A, 21, 369

\bibitem[{{Stix}(1976)}]{Stix76}
{Stix}, M. 1976, in IAU Symposium, Vol.~71, Basic Mechanisms of Solar Activity,
  ed. V.~{Bumba} \& J.~{Kleczek}, 367

\bibitem[{{Strugarek} {et~al.}(2018){Strugarek}, {Beaudoin}, {Charbonneau}, \&
  {Brun}}]{strugarek+18}
{Strugarek}, A., {Beaudoin}, P., {Charbonneau}, P., \& {Brun}, A.~S. 2018,
  \apj, 863, 35

\bibitem[{{Strugarek} {et~al.}(2017){Strugarek}, {Beaudoin}, {Charbonneau},
  {Brun}, \& {do Nascimento}}]{SBCBdN17}
{Strugarek}, A., {Beaudoin}, P., {Charbonneau}, P., {Brun}, A.~S., \& {do
  Nascimento}, J.-D. 2017, Science, 357, 185

\bibitem[{{Tayler}(1973)}]{tayler}
{Tayler}, R.~J. 1973, \mnras, 161, 365

\bibitem[{{Vidotto} {et~al.}(2014){Vidotto}, {Gregory}, {Jardine}, {Donati},
  {Petit}, {Morin}, {Folsom}, {Bouvier}, {Cameron}, {Hussain}, {Marsden},
  {Waite}, {Fares}, {Jeffers}, \& {do Nascimento}}]{Vidotto+14}
{Vidotto}, A.~A., {Gregory}, S.~G., {Jardine}, M., {et~al.} 2014, \mnras, 441,
  2361

\bibitem[{{Viviani} {et~al.}(2018){Viviani}, {Warnecke}, {K{\"a}pyl{\"a}},
  {K{\"a}pyl{\"a}}, {Olspert}, {Cole-Kodikara}, {Lehtinen}, \&
  {Brandenburg}}]{Viviani+18}
{Viviani}, M., {Warnecke}, J., {K{\"a}pyl{\"a}}, M.~J., {et~al.} 2018, \aap,
  616, A160

\bibitem[{{Vorontsov} {et~al.}(2002){Vorontsov}, {Christensen-Dalsgaard},
  {Schou}, {Strakhov}, \& {Thompson}}]{V+02}
{Vorontsov}, S.~V., {Christensen-Dalsgaard}, J., {Schou}, J., {Strakhov},
  V.~N., \& {Thompson}, M.~J. 2002, Science, 296, 101

\bibitem[{{Warnecke}(2018)}]{warnecke18}
{Warnecke}, J. 2018, \aap, 616, A72

\bibitem[{{Warnecke} {et~al.}(2014){Warnecke}, {K{\"a}pyl{\"a}},
  {K{\"a}pyl{\"a}}, \& {Brandenburg}}]{WKKB14}
{Warnecke}, J., {K{\"a}pyl{\"a}}, P.~J., {K{\"a}pyl{\"a}}, M.~J., \&
  {Brandenburg}, A. 2014, \apjl, 796, L12

\bibitem[{{Warnecke} {et~al.}(2018){Warnecke}, {Rheinhardt}, {Tuomisto},
  {K{\"a}pyl{\"a}}, {K{\"a}pyl{\"a}}, \& {Brandenburg}}]{Warnecke+18}
{Warnecke}, J., {Rheinhardt}, M., {Tuomisto}, S., {et~al.} 2018, \aap, 609, A51

\bibitem[{{Wright} \& {Drake}(2016)}]{Wright+16}
{Wright}, N.~J. \& {Drake}, J.~J. 2016, \nat, 535, 526

\bibitem[{{Wright} {et~al.}(2011){Wright}, {Drake}, {Mamajek}, \&
  {Henry}}]{Wright+11}
{Wright}, N.~J., {Drake}, J.~J., {Mamajek}, E.~E., \& {Henry}, G.~W. 2011,
  \apj, 743, 48

\bibitem[{{Yoshimura}(1975)}]{Y75}
{Yoshimura}, H. 1975, \apj, 201, 740

\bibitem[{{Zahn} {et~al.}(2007){Zahn}, {Brun}, \& {Mathis}}]{ZB07}
{Zahn}, J.-P., {Brun}, A.~S., \& {Mathis}, S. 2007, \aap, 474, 145

\bibitem[{{Zhao} {et~al.}(2013){Zhao}, {Bogart}, {Kosovichev}, {Duvall}, \&
  {Hartlep}}]{zbkdh13}
{Zhao}, J., {Bogart}, R.~S., {Kosovichev}, A.~G., {Duvall}, Jr., T.~L., \&
  {Hartlep}, T. 2013, \apjl, 774, L29

\end{thebibliography}

\begin{discussion}

\discuss{Brandenburg}{The cycle to rotation frequency is found
to  increase with stellar activity. Do you agree that there is currently no model that
reproduces this? How can we hope to reproduce the Sun if this essential feature
is not reproduced. Could you speculate what might be missing in the models?}

\discuss{Guerrero}{As a matter of fact no current model reproduces this relation. Nevertheless, 
cycle dynamos in global simulations are relatively new. There is still a broad
parameter space to explore in the simulations, specially concerning the resolution
and the energy transport issue. I'm optimistic that as soon as these issues are 
clarified the models will be able to reproduce better the solar and stellar 
magnetism. Note also that observations are still providing new results.}

\discuss{Strugarek}{The trend of differential rotation seems to increase
as rotation rate diminishes.  This is at odds with other 3D simulations with 
and without underlying stable layer. How is differential rotation sustained 
in your models?}

\discuss{Guerrero}{The latitudinal shear indeed increases when the rotational rate decreases. This
relation, not included in the presentation, is shown in Fig.~\ref{fig.os} of 
this proceeding. The results, however, are not completely at odds with other
simulations. See, for instance, the results presented in \cite{GYMRW14} including
global simulations from several groups. As rotation decreases the shear increases
before changing sing for anti-solar differential rotation.  In \cite{GSDKM16b}
we presented the angular momentum balance of one particular simulation, a detailed
analysis of all the simulations is still on its way. This will clarify the
reason for this relation.}

\discuss{Luhmann}{There is a well-known observed relationship between surface
polar fields and cycle size. Do the simulations reveal a physical reason for that
relationship?}

\discuss{Guerrero}{This property has not yet been studied in the results of global models. The reason
is that unfortunately there is not yet a model reproducing well all the properties
of the solar cycle. }

\end{discussion}

\end{document}